\documentclass
[preprint,nofootinbib,amsfonts,amssymb,bibnotes,onecolumn,12pt,prb]{revtex4}%
\usepackage{amsfonts}
\usepackage{amsmath}
\usepackage{amssymb}
\usepackage{graphicx}%
\providecommand{\U}[1]{\protect\rule{.1in}{.1in}}

\begin{document}
\preprint{UATP/09-07}
\title{Energy Gap and the Ideal Glass as a Defective Crystal: A Lattice Model of
Monatomic Systems}
\author{P. D. Gujrati}
\email{pdg@uakron.edu}
\affiliation{Departments of Physics and of Polymer Science, The University of Akron, Akron,
OH 44325}
\date{\today}

\begin{abstract}
We use the cell model to justify the use of a lattice model\ to study the
ideal glass transition. Based on empirical evidence and several previous exact
calculations, we hypothesize that there exists an \emph{energy gap} between
the lowest possible energy of a glass (the ideal glass IG) and the crystal
(CR). The gap is due to the presence of \emph{strongly correlated excitations}
with respect to the ideal CR; thus, one can treat IG as a highly defective
crystal. We argue that an excitation in IG requires energy that increases
logarithmically with the size of the system; as a consequence, we prove that
IG must emerge at a positive temperature $T_{\text{K}}.$ We propose an
antiferromagnetic Ising model on a lattice to model liquid-crystal transition
in a simple fluid or a binary mixture, which is then solved exactly on a
recursive (Husimi) lattice to investigate the ideal glass transition, the
nature of defects in the supercooled liquid and CR analytically, and the
effects of competing interactions on the glass transition. The calculation
establishes the gap. The lattice entropy of the supercooled liquid vanishes at
a positive temperature $T_{\text{K}}>0$, where IG emerges but where CR has a
positive entropy. The macrostate IG is in a particular and unique disordered
\emph{microstate} at $T_{\text{K}},$ just as the ideal CR is in a perfectly
ordered microstate at absolute zero. This explains why it is possible for CR
to have a higher entropy at $T_{\text{K}}$ than IG. The demonstration here of
an entropy crisis in monatomic systems along with previously known results
strongly suggests that the entropy crisis first noted by Kauzmann and
demonstrated by Gibbs and DiMarzio in long polymers appears to be ubiquitous
in all supercooled liquids.

\end{abstract}
\maketitle

\section{Introduction}

\subsection{Glass as a Stationary Metastable State (SMS)}

A common feature of almost all crystallizable materials is that they can
become glassy,\cite{Kauzmann} a state that is \emph{metastable.} \cite{Landau}
The true equilibrium state corresponds to its crystalline state (CR). The
glass is obtained by cooling the supercooled liquid (SCL). The metastable
states\emph{ }(MSs) have higher free energies than the corresponding
equilibrium state (EQS) and violate the fundamental thermodynamic property
that the equilibrium free energy be \emph{minimized} or the
equilibrium\ partition function (PF) be maximized.\emph{ }Even at absolute
zero, glasses are empirically found to have higher energies compared to their
crystalline form because of their metastability.\cite{Landau} We call this
difference in their energies the \emph{energy gap}. The existence of glasses
\emph{cannot} be rigorously demonstrated by \emph{equilibrium} statistical
mechanics, in which one considers all microstates in the PF. We need to go
beyond it to explain their existence. One way is to restrict the microstates
in the PF to be only disordered to describe glasses. As a consequence, many
standard results of equilibrium thermodynamics will not hold for metastable
states, \cite{Note-Continuation,Fisher,Suto,Domb0,Note01} even when they are
manipulated to exist for an abnormally long time, a situation that commonly
occurs for glasses.

Stable and abnormally long lasting MSs can be easily prepared; we only have to
recall the stability of medieval glasses. They also appear in many mean-field
theories including the van der Walls equation such as the Bragg-Williams
theory, \cite{Bragg} and in exact calculations. \cite{GujMeta,Gujun} They
emerge solely because we abandon the \emph{free energy minimization principle
}in the calculation. The violation of this principle can still lead to a
time-independent stable solution, except that it is metastable. The stability
only \ requires the specific heat, compressibility, etc. to be\emph{
}non-negative.\emph{ }Such mathematically stable time-independent solutions in
theoretical models imply that they will never decay. We therefore call
them\cite{GujMeta,Gujun} \emph{stationary} \emph{metastable states} (SMSs) to
distinguish them from MSs that are encountered in experiments and that usually
change with time. In experiments, MSs emerge by the very nature of preparing
the system such as a fast quench, after which they undergo structural
relaxation towards their respective SMS associated with the stable
solution.\cite{Gujrati-Relaxation} In this work, we will assume that there is
only one SMS at a given temperature and volume. A complete understanding of
time-dependent MSs is not possible without a comprehensive understanding of
corresponding SMSs. Thus, in this work, we focus our attention on
investigating SMSs only. This stable solution has the lowest free energy among
all metastable states. It is time-independent, and represents an "equilibrium
state" similar to the equilibrium liquid (EL), except that the former is
metastable; the true equilibrium state is CR. In the following, we will call a
SMS an equilibrium state, even if it is a metastable state. This should cause
no confusion. All other MSs represent time-dependent non-equilibrium states
(NEQSs). As the system relaxes from a NEQS towards SMS, the energy continues
to decrease. \cite{Gujrati-Relaxation} Thus, the energy gap of a SMS cannot be
larger than that for a NEQS.

There are usually two different mechanism operative in MSs. The
\textquotedblleft fast\textquotedblright\ mechanism (time scale $\tau
_{\text{f}}$) creates a metastable state in the system, followed by a
\textquotedblleft slow\textquotedblright\ mechanism (time scale $\tau
_{\text{s}}$) for nucleation of the stable phase and the eventual decay of the
metastable state. For an MS to exists for a while, we need to require
$\tau_{\text{s}}>\tau_{\text{f}}.$\ The time-dependent NEQSs include not only
states that will eventually turn into equilibrium states (such as crystals),
but also states that will eventually turn into SMSs (such as glasses) as we
wait infinitely long (in principle), depending on how they are prepared. To
study glass dynamics, we need to compare the two time scales with the longest
feasible experimental observation time $\tau_{\text{exp}}$. From the
experimental point of view, the inequality $\tau_{\text{s}}>>\tau_{\text{exp}%
}$ for supercooled liquids is almost equivalent to the long time, i.e. the
\emph{stationary} limit\ $\tau_{\text{s}}\rightarrow\infty$ of metastable
states$;$ we again appeal to the stability of medieval glasses. Thus, to a
first approximation, we can treat real glasses and SCLs as SMSs, states which
never decay. (According to Maxwell, \cite{Maxwell} this can, in principle, be
achieved by ensuring that the equilibrium state nuclei are absent in MSs.)
Above the glass transition, SCL is an equilibrium state in the sense of a SMS.
Thus, we will assume the existence of these SMSs, as they play a central role
in our modeling and understanding of glass formers; see the above discussion.
This allows us to use the PF formalism to study a SMS. The study of SMSs,
which can be carried out using the basic principles of statistical mechanical
formalism, though modified by imposing some \emph{restrictions }as described
later, still has a quite useful predictive value. The actual dynamics that
leads to such a state will not be our focus here, even though it is important
in its own right.

\subsection{Entropy Crisis}

The excess entropy (to be correctly replaced later by the communal entropy in
a continuum model or the entropy in a lattice model; see Sect.
\ref{Lattice_Model}) of SCL exhibits a rapid drop below the melting
temperature $T_{\text{M}},$ \cite{Kauzmann} and eventually vanishes at some
temperature $T=$ $T_{\text{K}}<T_{\text{M}}.$ It will become negative if
extrapolated to lower temperatures. A negative communal entropy is considered
\emph{unphysical}. Thus, the extrapolation gives rise to an \emph{entropy
crisis} at $T_{\text{K}}$. An ideal glass transition is invoked to avoid this
crisis at $T_{\text{K}}$. In the limit of zero cooling rate (not accessible in
experiments or simulations, but accessible in a theoretical setup) in the
metastable region, metastable states are nothing but SMSs, and can be
described by "equilibrium" statistical mechanics by restricting the allowed
microstates to be disordered; microstates leading to the crystalline state are
not allowed. In experiments, the extrapolated excess entropy of many glassy
materials at $T=0$ \cite{GujGld} is found to have a non-zero value depending
on the rate of cooling. This does not rule out the possibility that the
entropy of the hypothetical "stationary" glass vanishes in the limit of zero
cooling rate at a positive temperature. \cite{note00} Furthermore, the amount
of entropy discontinuity at melting is not important for the existence of an
ideal glass transition; all that is required is the existence of a supercooled
liquid relative to a CR in which the entropy is very small. The latter can
happen either by a very large latent heat or by a rapid drop in its entropy;
see the behavior of CR entropy in Figs. \ref{F1} and \ref{F2}.

Demonstrating an entropy crisis for supercooled liquids in a restricted
formalism has been one of the most challenging problems in theoretical physics
because of the inherent approximation involved in most calculations. An
entropy crisis in long polymers was theoretically demonstrated almost fifty
years ago by Gibbs and DiMarzio \cite{GibbsDiMarzio,GujGold,GujC,GujRC} to
support the entropy crisis as a \emph{fundamental principle} underlying glass
transitions in long polymers. Their work was later criticized by Gujrati and
Goldstein \cite{GujGold} for its poor approximation, and doubt was cast on
whether the entropy crisis in long polymers was genuine. The situation changed
when in a recent work, Gujrati and Corsi \cite{GujC,GujRC} established the
existence of an entropy crisis in long polymers by using a highly reliable
approximate approach. This was very important as the idea of Gibbs and
DiMarzio has been pivotal in shaping our thinking about the ideal glass
transition. Recently, we have also succeeded in demonstrating the entropy
crisis in a dimer model \cite{GujS,Fedor} containing anisotropic interactions.

\subsection{Current Goal}

Our goal in the present work is two-fold. First, we demonstrate the crisis in
simple isotropic fluids containing monatomic particles. The second aim is to
understand the nature of a general glassy state by investigating monatomic
glass formers and the lessons learned from our previous
investigations.\cite{GujGold,GujC,GujRC,Fedor,GujS,GujMeta,Gujun} Study of
monatomic glass formers also allow us to obtain a better understanding of the
glassy structure (defects therein) whose accurate representation still remains
challenging. For the general investigation, we propose a generic model of the
entropy, which is consistent with our previous exact
calculations\cite{GujGold,GujC,GujRC,Fedor,GujS,GujMeta,Gujun} and the
calculation presented here for monatomic glass formers. The slow relaxation
\cite{Kauzmann} in SCLs is similar to that observed in ordinary spin glasses,
\cite{Morgenstern} whose important features are their geometrical frustration
and competition. We will, therefore, also investigate whether the competition
and frustration play an important role in promoting the glassy behavior in
monatomic glass formers.

It should be noted that frustrated antiferromagnets (AF) and spin glasses do
not usually possess long range order at low temperatures because of \ a highly
degenerate ground state \cite{Ramirez} and their glassy behavior is brought
about by the presence of frustration or quenched impurities and is somewhat
well understood. In contrast, supercooled liquids require a \emph{unique}
ground state, the crystal. This distinguishes the glassy behavior in
supercooled liquids and requires considering an \emph{unfrustrated} AF model
as a paradigm of simple fluids or alloys in this work. We consider a pure
\ (no frustration or quenched impurities) AF Ising model, which possesses a
unique ordered state, which is identified with CR, so that supercooling can
occur. This then results in a glassy state. The use of a magnetic model to
study glassiness is highly desirable as magnetic systems have been extensively
studied in theoretical physics and are well understood at present. This makes
interpreting the results very easy and transparent. We are not aware of any
simple model calculation to date to justify glassy states in a pure AF model.
The model allows us to investigate the defects that occur in the ideal glass
(with respect to the corresponding crystal). We also find that the competition
considered in this work inhibits instead of promoting the glass transition,
which is a surprising result.

\subsection{Layout}

The layout of the paper is as follows. In the next section, we use the
celebrated cell model of fluids to argue that it is the communal entropy that
should determine the location of the glass transition, in which the glass
"melts" from its localized state into a delocalized liquid state. It is shown
that a lattice model is sufficient to determine the glass transition. The
thermodynamics of metastability, the role of the energy gap on the entropies
of ordered and disordered states, and the possible free energies of the two
states are discussed in Sect. \ref{Singularity_Assumption}. The conjectural
form of the entropy is based on the work of Gibbs and
DiMarzio\cite{GibbsDiMarzio} and exact calculations from our group.
\cite{GujGold,GujC,GujRC,Fedor,GujS,GujMeta,Gujun} The lattice model is
introduced in Sect. \ref{Model}, and is analytically solved in Sect.
\ref{Binary_Mixture}. The results are presented in Sect. \ref{Results}. This
section forms the core of the present work and contains important results
including the demonstration of an energy gap, the existence of an ideal glass
transition, the discussion of defects and the nature of the ideal glass as a
single disordered microstate. The last section contains a brief summary of the work.

\section{Cell Model\label{Cell_Model0}}

\subsection{Localization or Confinement\label{Cell_Model}}

One of the most important property of SCLs at low temperatures is that glasses
and CRs have very similar vibrational heat capacities below $T_{\text{g}}$,
except that glasses have higher potential energies than the corresponding CRs.
\cite{Kauzmann,SimhaGoldstein,Landau} Let $E_{\text{NEQS}}$ denote the lowest
possible energy of a NEQS such as a glass and $E_{0}$ the energy of the ideal
crystal at $T=0$. Then, we empirically have
\begin{equation}
E_{\text{NEQS}}>E_{0}. \label{Enery_Diff_0}%
\end{equation}
The difference%
\begin{equation}
\Delta_{\text{G}}\equiv E_{\text{NEQS}}-E_{0}>0 \label{Energy_Gap}%
\end{equation}
is called the \emph{energy gap}. Otherwise, glasses and crystals are confined
to execute quite similar vibrations (not necessarily harmonic) within their
potential wells\emph{ }or\emph{ }basins, although their minima are at
different energies, to a first approximation, notwithstanding the clear
evidence to the contrary. \cite{Goldstein-ExcessEntropy} All that is important
is that both glasses and crystals exhibit localized motion. This property has
led to the enormous popularity of the \emph{potential energy landscape}
picture, originally proposed by Goldstein, \cite{Goldstein} to investigate
glass transition in a system consisting of $N$ particles. The potential energy
$E(\{\mathbf{r}\})$ as a function of the set $\{\mathbf{r}\}$ of particles'
positions uniquely determines this landscape. A glass is \emph{confined} to
one of the myriads of potential wells, while the crystal is most commonly
believed to be confined to just one potential well. The number of these basins
determines the basin entropy, and is a measure of the entropy, which Goldstein
identifies as the residual entropy. \cite{Goldstein-ExcessEntropy}

\begin{figure}
[ptb]
\begin{center}
\includegraphics[
trim=0.000000in 7.814348in 3.728873in 0.000000in,
height=2.8496in,
width=4.2661in
]%
{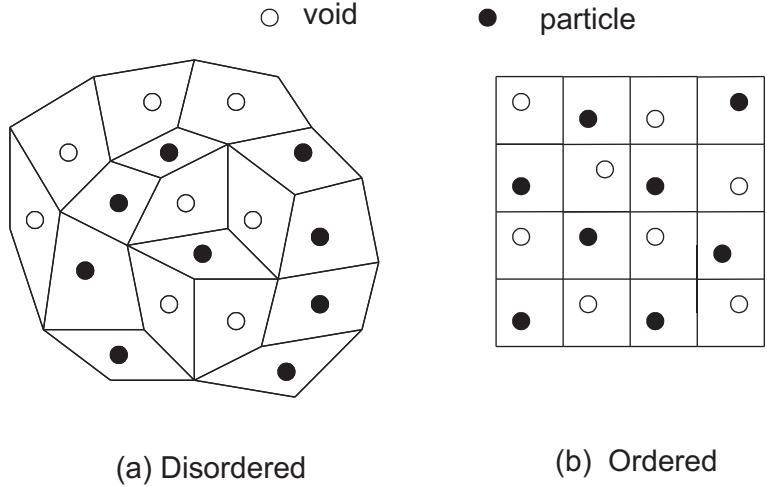}%
\caption{Cell representation of a small region of disordered (a) and ordered
(b) configurations at half occupation: only half of the cells contain a
particle each. Observe that while each particle is surrounded by four voids in
the ordered configuration, this is not the case for the disordered
configuration. We have shown a higher volume for the disordered configuration,
as found empirically. }%
\label{Fig_Random_Ordered_States}%
\end{center}
\end{figure}

A glass is SCL trapped in one of the many basins at $T_{\text{g}},$ and
executes vibrations within this potential well. The resulting glass can be
characterized by the properties of this basin, \cite{Kauzmann,Goldstein} which
determine the \emph{average} configuration of the glass in that basin. The
high barriers of these potential wells also provide high stability to very
slowly cooled SCLs, at least near $T_{\text{g}}$, and are responsible for the
strong inequality $\tau_{\text{s}}>>\tau_{\text{exp}}.$ This also implies that
barriers to the formation of stable nuclei must also be extremely high in
SCLs. This was first argued by Goldstein in his seminal work on viscous
liquids. \cite{Goldstein}
\begin{figure}
[ptb]
\begin{center}
\includegraphics[
trim=0.000000in 7.822857in 3.724959in 0.000000in,
height=2.8548in,
width=4.2687in
]%
{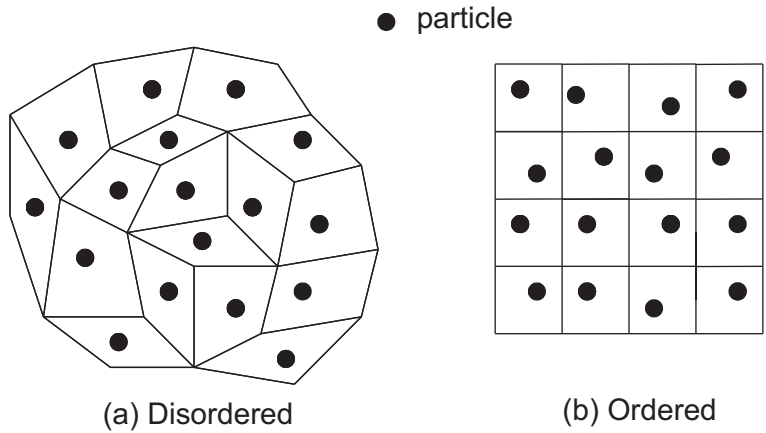}%
\caption{Cell representation of a small region of disordered (a) and ordered
(b) configurations at full occupation: each cell contains a particle. Each
cell representation uniquely defines a potential well or basin in the
potential energy landscape. Observe that while each particle is surrounded by
four particles in the ordered configuration, this is not the case for the
disordered configuration. We have shown a higher volume for the disordered
configuration, as found empirically; see later. }%
\label{Fig_Random_Ordered_States0}%
\end{center}
\end{figure}
This picture of slowly-cooled SCLs allows us consider the liquid cell model of
Lennard-Jones and Devonshire, \cite{Lennard-JonesDevonshire} see Fig.
\ref{Fig_Random_Ordered_States0}, and its various elaborations.
\cite{Kirkwood} We show a cell representation of a dense liquid in (a) and of
a crystal in (b). Each cell is occupied by a particle in which it vibrates.
The regular lattice in (b) is in accordance with Einstein's model of a
crystal. In the liquid state, this regularity is absent. A defect in the cell
representation corresponds to some empty cells, as shown in Fig.
\ref{Fig_Random_Ordered_States}. We consider the (configurational) partition
function $Z(T,V)$\
\begin{equation}
Z(T,V)\equiv\frac{1}{v_{0}^{N}N!}\int e^{-\beta E}d^{N}\{\mathbf{r\}}%
\equiv\int W(E,V)e^{-\beta E}dE/\epsilon_{0}\mathbf{,} \label{CPF}%
\end{equation}
in the canonical ensemble at temperature $T$ (measured in the units of the
Boltzmann constant $k_{\text{B}};$ this amounts to effectively setting
$k_{\text{B}}=1$)$;$ $\beta\equiv1/T$ is the inverse temperature, and
$W(E,V)dE/\epsilon_{0}$ represents the number of distinct
configurations\cite{NoteEntropy} with potential energy in the range $E$ and
$E+dE$; $v_{0}$ and $\epsilon_{0}$\ represent some small-scale constant volume
such as the cell volume, and some energy constant, such as the average spacing
between vibrational energy levels of a single particle in its cell in Fig.
\ref{Fig_Random_Ordered_States0}(b) at $T=0$. We set $v_{0}=1$ and
$\epsilon_{0}=1$\ in this work.

One should carefully make a distinction between a quantity and its
configurational component by adding a subscript "T' to the quantity. For
example, $S_{\text{T}}(T,V)$ will denote the (total) entropy, while
$S(T,V)$\ the configurational part of $S_{\text{T}}(T,V)$. The
\emph{microcanonical} configurational entropy is given by by the Boltzmann
relation so that\cite{note0}
\[
S(E,V)\equiv\ln W(E,V)dE/\epsilon_{0}\simeq\ln W(E,V)\geq0.
\]
The average energy $\overline{E}=E(T,V)$ is used to give the \emph{canonical}
configurational entropy $S(T,V)\equiv S[E(T,V),V].$ The slope of the entropy
at $\overline{E}$ is related to the inverse temperature by the standard
relation valid at equilibrium
\begin{equation}
\left.  (\partial S/\partial E)_{V\ }\right\vert _{\overline{E}}=\beta.
\label{Inverse_Temp}%
\end{equation}

This entropy differs from the total entropy $S_{\text{T}}(T,V)$, which is the
entropy associated with the total PF $Z_{\text{T}}(T,V)$ where the kinetic
energy (KE) is also included. \cite{GujRC,GujFedor,Fedor,GujS} The entropy
contribution $S_{\text{KE}}(T)$ due to the kinetic energy is the \emph{same}
for all systems as it is independent of the interactions and volume. For this
reason, we do not have to include $S_{\text{KE}}(T)$ in any investigation of
metastability. We only consider the \emph{configurational degrees of freedom}
from now on. For the same reason, there is also no need to explicitly show the
dependence on $V$. The configurational part of the Helmholtz free energy is
given by $F(T)\equiv-T\ln Z(T)$, where we have suppressed $V.$

\subsection{Communal Entropy\label{Lattice_Model}}

The communal entropy is defined as the difference between the configurational
entropy $S(T)$ of the system and the entropy $S_{\text{b}}(T)$ when the
particles are \emph{confined} in their cells, \cite{Hill} i.e., the basin (we
suppress showing the $V$-dependence):%
\begin{equation}
S_{\text{comm}}(T)\equiv S(T)-S_{\text{b}}(T)\geq0. \label{Communal_S}%
\end{equation}
The communal part of the free energy is given by%
\begin{equation}
F_{\text{comm}}(T)\equiv F(T)+TS_{\text{b}}(T)=E(T)-TS_{\text{comm}}(T).
\label{Communal_FreeEnergy}%
\end{equation}
\emph{The communal entropy is the entropy due to the deconfinement of the
system from the basin. }It is a measure of the diffusional contribution to the
total entropy as opposed to the vibrational entropy. Accordingly, it vanishes
when the system is confined in a basin such as in Fig.
\ref{Fig_Random_Ordered_States0}(a). Let $T_{\text{K}}>0$\ denote the
temperature when this happens for SCL:%
\begin{equation}
S_{\text{comm}}^{\text{SCL}}(T_{\text{K}})\equiv0. \label{S_comm_Ideal_Glass}%
\end{equation}
Several authors identify the communal entropy as the configurational entropy.
We, however, reserve this name for the entropy associated with the
configurational partition function (\ref{CPF}). Another appropriate term for
the communal entropy is the residual entropy. Here, we will stick with the
term communal entropy, which conveys directly the idea of delocalization. From
now on, we will simply call any configurational quantity $Q(T)$ such as
$S(T)$\ as the quantity; we will refer to $Q_{\text{T}}(T)$ as the total quantity.

As noted in Sect. \ref{Cell_Model0}, it is a common assumption that CR is also
confined to a single basin such as the one shown in Fig.
\ref{Fig_Random_Ordered_States0}(b). If the assumption holds, then the
confinement will occur at some $T_{\text{CR}}>0$:%
\begin{equation}
S_{\text{comm}}^{\text{CR}}(T_{\text{CR}})\equiv0. \label{S_comm_CR}%
\end{equation}
As we will see later in Sect. \ref{Results}, this does not happen due to the
presence of defects that emerge in CR. Thus, the CR lattice entropy (the
communal entropy) vanishes only at absolute zero. We will also see there that
the defects in IG are strongly correlated, making them different in nature
from those in CR. We will discuss this issue further in Sects.
\ref{Singularity_Assumption} and \ref{Results}.

In confined or localized states, particles only occupy positions that are
within their individual cells. Let us focus on a SCL localized into one such
basin. Let the minimum of the basin energy be $E_{\text{K}}$. It corresponds
to the average state in the basin in which each particle has its average
position. Any deviation from these positions will only raise the potential
energy. The relevant average state in the basin at $T=T_{\text{K}}$ remains
unchanged below $T_{\text{K}}$ and continues to represent the average state of
the system. In other words, the average state of the system remains frozen in
this \emph{inert average state} below $T_{\text{K}}$ and is called the\emph{
ideal glass}$\ $(IG)$.$ This localization and freezing of the average state to
IG is called the ideal glass transition. Thus, the vanishing of
$S_{\text{comm}}(T)$ in (\ref{S_comm_Ideal_Glass}) is taken as the condition
for the formation of an IG due to its localization or confinement in a single
basin, and the temperature $T_{\text{K}}$ at which this occurs is called the
ideal glass transition temperature or the Kauzmann temperature, to honor the
pioneering contribution of Kauzmann in the field of supercooled liquid, even
though the term is commonly used to denote the temperature where the CR and
SCL have the same entropy. This issue has been discussed elsewhere,
\cite{GujFedor,Fedor,GujS} to which we refer the reader for additional information.

\begin{quote}
\textsc{Remark} \textit{In a lattice model, particles do not deviate from
their fixed lattice positions. Consequently, the total entropy in the lattice
model is purely communal in nature}$.$
\end{quote}

Thus, a lattice model can be effectively used to analyze the average state of
a glass former in real continuum space. Accordingly, we do not feel guilty
about using lattice models here for which exact calculations can be carried
out. As the entropy of the IG is zero, \cite{note0} it means that the
macrostate corresponding to IG is really a microstate. In other words, the IG
at $T_{\text{K}}$ refers to a particular disordered microstate of energy
$E_{\text{K}}$ in a lattice model. This is no different from the ideal CR at
absolute zero, which is also in a the perfectly ordered microstate.

Since the heat capacity is non-negative, $S(T),$ and $E(T)$ are monotonic
increasing function of $T$, and must have their minimum values at absolute
zero. Assuming CR to be the stable phase at absolute zero, we conclude that it
must be in the state with the lowest possible energy $E_{0}(V).$ This follows
from the Nernst-Planck postulate \cite{NernstPostulate}
\[
TS_{\text{CR}}(T,V)\rightarrow0\ \text{as }T\rightarrow0,
\]
so that
\[
F(T=0,V)=E_{0}(V).
\]
As long as the heat capacities of various phases, stable or metastable, remain
non-negative, and we will see that this is true, all higher energies will
correspond to temperatures $T>0$.\textbf{ }As the lowest possible energy of
the glass is $E_{\text{K}}>E_{0}$, this energy gap will suggest that the SCL
will get into the IG microstate at $T_{\text{K}}>0.$ We have seen it to be
true in all of our exact calculations so far;
\cite{GujC,GujRC,Fedor,GujS,GujMeta,Gujun} it is also true in the present
calculation, as we will see later in Sect. \ref{Results}. The calculation by
Gibbs and DiMarzio \cite{GibbsDiMarzio} is also consistent with this claim.

\section{Thermodynamics of Metastability\label{Singularity_Assumption}}

\subsection{Schematic Communal Entropy}

\begin{figure}
[ptb]
\begin{center}
\includegraphics[
trim=0.000000in 2.839152in 0.000000in 1.838851in,
height=3.6106in,
width=4.9216in
]%
{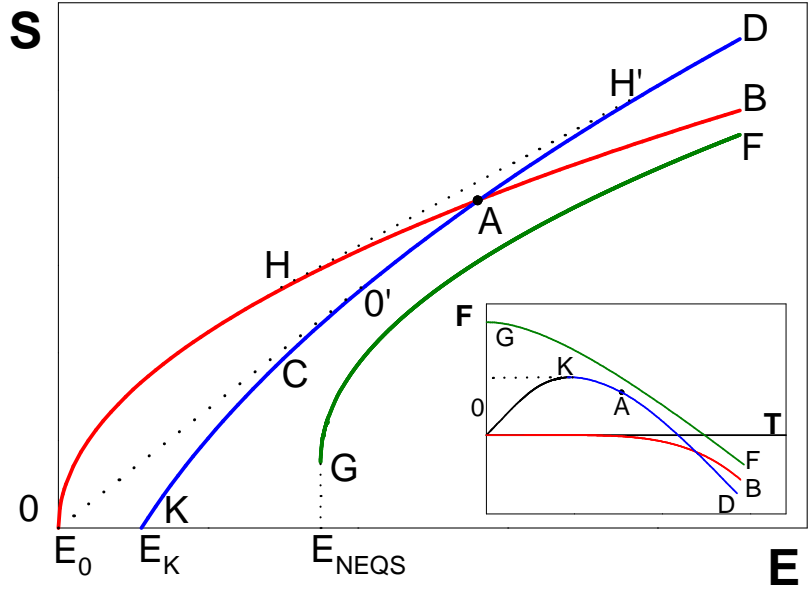}%
\caption{Schematic form of generic communal entropy functions defined in
(\ref{Communal_S}) as continuous and \emph{concave} functions \cite{concave}
of $E$ for a fixed volume $V$ for various possible states, and the resulting
continuous and concave Helmholtz free energies in the inset. Note the presence
of an energy gap ($\Delta_{\text{G}}>0$). This form, which is shown in an
exaggerated fashion to highlight the distinction, will be justified by our
results in Sect. \ref{Results}. The form is also consistent with all known
exact calculations from our group. \cite{GujC,GujRC,GujS,Fedor,CorsiThesis}The
point O$^{\prime}$ represents the point where the free energy DO$^{\prime}$CKO
of the liquid in the inset is equal to the free energy of the crystal at O
(absolute zero). The point A on DO$^{\prime}$CKO in the inset is slightly
below the melting temperature $T_{\text{M}},$ where it crosses the crystal
free energy OAB. }%
\label{Fig-S-T}%
\end{center}
\end{figure}

There are no general arguments \cite{Landau,Ruelle,Huang} to show that
thermodynamically stable states must always be ordered, i.e., periodic. The
remarkable aperiodic Penrose tilings\ of the plane, for example, by two
differently but suitably shaped tiles are stable. It is found empirically that
the volume (or the energy or enthalpy) of a glass, or more generally, an NEQS
at absolute zero is higher than that of the corresponding crystal;
see\cite{GujGld} for example for a careful analysis of data. The energy gap is
also seen in the original calculation of Gibbs and DiMarzio,
\cite{GibbsDiMarzio} and many exact calculations carried out in our group.
\cite{GujGold,GujC,GujRC,Fedor,GujS,GujMeta,Gujun} Here, we will focus on the
potential energy for which the above observation is in accordance with
(\ref{Enery_Diff_0}); see Fig. \ref{Fig-S-T}. (There will be no energy gap if
$E_{\text{NEQS}}=E_{0}.$) According to (\ref{Enery_Diff_0}), there must be
many \emph{defects} in the glass relative to the crystal even at absolute zero
to account for this difference in the energy (or enthalpy). The number and
nature of these defects must ensure that the glassy state is not only
disordered but also has a very small amount of communal entropy. The value of
$E_{\text{NEQS}}$ depends on the rate of cooling $r$. As $r$ decreases, this
energy falls and approaches a limiting value $E_{\text{K}}\leq E_{\text{NEQS}%
},$ which is still higher than $E_{0}$; see Sect. \ref{Binary_Mixture}. As CR
is heated, its energy rises due to the defects, but their densities is much
smaller in the crystal to leave it ordered; compare it with the corresponding
glass at the same temperature, which is disordered and has much higher energy.
As a consequence, one can treat a glass as a highly defective crystal
\cite{Gujrati-Glass-Defective-Crystal} with so many defects to the point that
the crystal becomes disordered.%

To form a glass, crystallization must be avoided either by ensuring that the
material does not have time to become crystalline or by suppressing the
mechanism to form a crystal. As glasses, or more generally NEQSs, are formed
under some sort of constraints (crystallization is forbidden in the case of
glasses), their (communal) multiplicity $W_{\text{NE}}(E)$ (which may be a
function of time, but we do not show it) must not be greater than
$W_{\text{EQ}}(E)$ of the corresponding equilibrium state, so that we must
always have\
\begin{equation}
S_{\text{NE}}(E)\leq S_{\text{EQ }}(E); \label{Second_Law_Entropy}%
\end{equation}
see Fig. \ref{Fig-S-T} for fixed $V$. The curve OHH$^{\prime}$D (with straight
segment HH$^{\prime}$) represents $S_{\text{EQ }}(E)$ and the curve GF
represents $S_{\text{NE}}(E).$ In time, the curve GF\ will move upwards
towards OHH$^{\prime}$D: it either converges to it if crystallization occurs,
or to KCAD if it is forbidden. This is consistent with the second law of
thermodynamics: As the constraints are removed, the entropy of a closed system
(fixed $E,V$) \emph{cannot} decrease in time; it can only increase or remain
constant. It is interesting to note that this entropy form is consistent with
the calculation of Gibbs and DiMarzio. \cite{GibbsDiMarzio}

From now on, we will only consider a lattice model in which the entropy is the
communal entropy shown schematically in Fig. \ref{Fig-S-T}. The segment
H$^{\prime}$D represents the entropy of the liquid, the disordered phase, and
is determined by the multiplicity $W_{\text{dis}}(E)$ of disordered
configurations. The segment OH represents the entropy of the crystal, the
ordered phase, and the entropy is determined by the multiplicity
$W_{\text{ord}}(E)$ of ordered configurations. Because of the straight segment
HH$^{\prime},$ the equilibrium entropy $S_{\text{EQ }}(E)$ is a singular
function, which is then reflected in a singular equilibrium free energy at the
melting temperature $T_{\text{M}}$; the latter is given by the inverse of the
slope of HH$^{\prime}$, see (\ref{Inverse_Temp})$.$ (It should be stressed
that the actual value of the energy discontinuity at melting, reflected by the
tangent construction HH$^{\prime}$, is not relevant for the investigation of
metastability.) Let $E_{\text{CR,M}},$ and $E_{\text{EL,M}}$ denote the
energies of the coexisting phases CR\ and EL at $T_{\text{M}}$;\cite{Note01}
see points H and H$^{^{\prime}}$ in Fig. \ref{Fig-S-T}$.$ It is hard to
imagine that the ordered and disordered configurations terminate at
$E_{\text{CR,M}}$ and $E_{\text{EL,M}},$ respectively. Thus, we will assume
that the curve KCAH$^{\prime}$D represents the entropy
\[
S_{\text{dis}}(E)\equiv\ln W_{\text{dis}}(E)
\]
of the disordered states in the system even below $E_{\text{EL,M}}$.
Similarly, we will assume OHAB\ to represent the entropy
\[
S_{\text{ord}}(E)\equiv\ln W_{\text{ord}}(E)
\]
of the ordered states in the system even above $E_{\text{CR,M}}$. The entropy
\ then has two different branches KCAD and OHAB, rather than a single function
given by OHH$^{\prime}$D. We only require that the branches be continuous and
concave. \cite{concave} The existence of the two branches requires that we are
able to distinguish between disordered and ordered states. After all, the
glassy state is formed by disordered states. So, making such a distinction is
not merely an academic curiosity; it is of vital importance for our
understanding of how and why glasses are formed.

\subsection{Non-zero $T_{\text{K}}$\label{Non-zero TK}}

As we wait longer and longer, NEQS will approach its two possible stationary
limits: EQS or SMS. In the process, the energy $E_{\text{NEQS}}$ will continue
to decrease.\cite{Gujrati-Relaxation} It will converge to $E_{0}$ if NEQS
approaches CR and there will be no energy gap; otherwise, it will converge to
some higher energy $E_{\text{K}}$ with an energy gap, if NEQS converges to
IG/SCL. The presence of the gap will turn out to imply \cite{Gujun} a non-zero
$T_{\text{K}}>0$. The following provides another and more illustrative proof
of this claim for the point K in Fig. \ref{Fig-S-T}. The energy gap requires a
large amount of defects to give rise to a disordered state associated with K,
where the entropy also vanishes. The defects at K must be very special in that
they cannot be uncorrelated like point-like defects that appear in CR. They
must be strongly correlated not only to make the particular IG microstate
disordered, but also to ensure that one cannot obtain other disordered
microstates of the same energy by simply exchanging distinct defects. The
latter condition must hold for IG to have zero entropy. Uncorrelated distinct
defects can be exchanged without changing the energy to generate many distinct
microstates. In the process we raise the entropy at K, which contradicts that
the entropy vanishes there. Thus, we conclude that the defects in IG are
strongly correlated.

Any defects, whether local (finite in size or independent of the lattice size
$N$) or non-local\cite{Note1} (growing with the lattice size $N$ so that it
becomes infinitely large in the limit $N\rightarrow\infty$), each can be
created in in $\sim N$ different ways by simply moving its "center-of-mass" to
any of the $N$ lattice sites. One may also be able to rotate the defect
without affecting the energy. Thus, the gain in the entropy is
\begin{equation}
\left.  \Delta S\right\vert _{V}\sim\ln N.\label{Entropy_Gain}%
\end{equation}
This remains true whether we consider an ordered state or a disordered state.

Let us now follow the consequence of the defects. The uncorrelated point-like
defects in a perfect CR are local defects that each require a small
non-extensive energy of excitation without destroying the order completely.
Using (\ref{Entropy_Gain}) for a defect, we have for their ratio%
\begin{equation}
\left(  \frac{\Delta E}{\Delta S}\right)  _{V}\sim1/\ln N,\label{CR_Temp}%
\end{equation}
which vanishes as $N\rightarrow\infty$. This explains, see (\ref{Inverse_Temp}%
), why the corresponding temperature of the perfect CR is $T=0$. On the other
hand, the defects in a glass are highly correlated. If these defects were of
finite size, they require a finite amount of energy of excitation. Then the
above argument applied to CR will suggest not only that IG can only emerge at
absolute zero, but also that one can also lower the energy of IG by removing
some of these defects. Lowering of the energy has two very important
consequences. The first one violates the condition that $E_{\text{K}}$ is the
lowest energy of IG. The second one is that removal of the defects can be done
in many ways to make the entropy of the lower energy state non-zero, which
then violates the concavity condition.\cite{concave} Thus, we conclude that
the defects in IG are not finite in size. Accordingly, their creation must
raise the energy that cannot remain finite. Let us assume that the excitation
energy $\propto N^{p}$ for a defect, where $1\geq p>0$. Using
(\ref{Entropy_Gain}) for a defect, we find the ratio
\begin{equation}
\left(  \frac{\Delta E}{\Delta S}\right)  _{V}\sim\frac{N^{P}}{\ln
N}\rightarrow\infty,\label{Glass_Temp0}%
\end{equation}
as $N\rightarrow\infty$ provided $p>0$. From (\ref{Inverse_Temp}), this will
then correspond to an infinite temperature $T_{\text{K}}$, which is certainly
not a physically relevant situation. The only meaningful way one can have
non-localized defects in IG is to have the excitation energy due to a defect
to be%
\begin{equation}
\left.  \Delta E\right\vert _{V}\sim\ln N.\label{IG_Excitation}%
\end{equation}
In this case and using (\ref{Entropy_Gain}), the ratio%
\begin{equation}
\left(  \frac{\Delta E}{\Delta S}\right)  _{V}=\text{ constant }%
>0\label{Glass_Temp}%
\end{equation}
as $N\rightarrow\infty$. It now follows immediately that the IG transition
occurs at a non-zero and finite temperature, which we identify with
$T_{\text{K}}$. This proves our claim.

It should be emphasized that the proof of the claim is based on the existence
of the energy gap. Thus, its validity provides a verification of our gap
model. The claim turns out to be consistent with the results in Sect.
\ref{Results}. It is also consistent with our previous exact results.
\cite{GujC,GujRC,GujS,Fedor,CorsiThesis} The following thermodynamic argument
also supports the claim. The vanishing of the entropy at K means that the free
energy has a maximum there at $T_{\text{K}}$. Thus, except in some special
cases, the free energy is quadratic in $T$ at $T_{\text{K}}$ in the leading
order:%
\[
F(T)=F(T_{\text{K}})-a(T-T_{\text{K}})^{2}+\cdots,
\]
where $a>0$ is a constant in $T$. Thus, both the entropy and the excitation
energy are linear in ($T-T_{\text{K}})$ in the leading order, so that they are
proportional as required by (\ref{Glass_Temp}). All these verifications
provide a strong validation of the energy gap model. As the slope of DAK in
Fig. \ref{Fig-S-T} is finite at K, this allows the possibility of continuing
DAK to the region below $E_{\text{K}}$, where one will encounter a negative
communal entropy. This possibility of the extension of DAK below K is why
various calculations give this unphysical
branch.\cite{GibbsDiMarzio,GujC,GujRC,GujS,Fedor,CorsiThesis}

The strong correlations produced by these non-localized defects in IG must be
of such a nature that one cannot reduce the energy of the IG microstate below
$E_{\text{K}}$. How does this happen can only be answered after we can
identify the IG microstate, which at present is an unsolved problem. It is
hoped that the present simple but analytical solution presented in Sect.
\ref{Binary_Mixture} will be a first step towards this goal. It should also be
said at this point that the above discussion for IG is not applicable to real
glasses described by the curve FG in Fig. \ref{Fig-S-T}. The central point in
the above discussion was the impossibility of generating disordered
microstates of energies lower than $E_{\text{K}}$. However, this is not
applicable at the point G, since one can certainly produce disordered
microstates of energies lower than $E_{\text{NEQS}}$. As a consequence, the
defects in real glasses are going to be local and not non-local. Accordingly,
the temperature at G will be zero; see (\ref{CR_Temp}). Even the entropy there
does not have to vanish. 

\subsection{Order Parameter\label{Order_Parameter}}

We have already argued elsewhere\cite{GujS} that the ideal glass transition is
a continuous transition at which the free energy will exhibit a thermodynamic
singularity. The result is based on the form of entropy shown in Fig.
\ref{Fig-S-T}. Therefore, it is necessary to identify an order parameter for
the IG transition. For this, we turn to the melting of crystals into liquids,
which is similar (but not identical) to the "melting" of glasses into SCLs.
This will require leaving the lattice model for a while. According to the
Lindemann criterion of melting, \cite{Lindemann} a crystal melts when the
mean-square displacement $\overline{\mathbf{r}^{2}}$ becomes so large that the
atoms start to get into each other's cells to the point that they begin to
diffuse over a large distance, and the melting initiates. If $a$ denotes the
interatomic distance in the crystal, then the melting proceeds when
$\overline{\mathbf{r}^{2}}=c_{\text{L}}a^{2},$ where $c_{\text{L}}$ is
Lindemann's constant and is expected to be the same for crystals with similar
structure. It should, however, be pointed out that the criterion is expected
to be valid only for those systems that have simple crystalline structures.
For particles obeying Lennard-Jones potential, Jin and coworkers \cite{Jin}
have tested the validity of the criterion by considering 6912 Lennard-Jones
particles. The value of $c_{\text{L}}$ is estimated to be $\simeq0.12$-$0.13$
at equilibrium melting, as expected. As the temperature is raised towards the
melting temperature, clusters of correlated particles of various sizes are
formed. At the melting, the correlations become too strong and the density of
defects becomes large enough that the crystal melts. These defects in CR are
necessary to melt it and require a macroscopic energy $\Delta E\propto N$ to
be created, which is reflected in the energy discontinuity. The associated
entropy discontinuity $\Delta S\propto N$\ is such that the ratio%
\[
\left(  \frac{\Delta E}{\Delta S}\right)  _{V}=T\,_{\text{M}},
\]
compare with (\ref{Glass_Temp}), which follows from the equality of the free
energy $F$ in the two phases in coexistence. This equation should be compared
with (\ref{Inverse_Temp}). The equality of the two follows from the linear or
tangent construction which gives rise to HH$^{^{\prime}}$ in Fig.
\ref{Fig-S-T}. If we now treat a glass as a defective crystal
\cite{Gujrati-Glass-Defective-Crystal} with a disordered cell representation,
then it is not hard to imagine that a similar criterion can be applied to
the"melting" of a glass into its SCL state. We will pursue this analogy a bit
further below.

The so-called melting of an ideal glass into a SCL is nothing but a
localization-delocalization transition, as discussed in Sect. \ref{Cell_Model}%
, similar to the melting transition. As the temperature is raised, the IG
becomes deconfined and begins to probe via diffusion other basins
corresponding to different cell representations, thereby giving rise to the
$\alpha$ relaxation. We can make this process somewhat
quantitative\cite{Novikov} by considering a particle at some average position
$\mathbf{r}_{0}$ within its cell. We average the square of its displacement.
At low temperatures where it is confined within its cell, the particle does
not diffuse away from the original point $\mathbf{r}_{0};$ it merely undergoes
vibrations within its cell. Let us introduce a certain length $b$
characterizing the average size of the cell. Then, $\overline{\mathbf{r}^{2}%
}\leq c_{\text{G}}b,$ with $c_{\text{G}}<1$ a parameter that depends on the
system under investigation, will characterize a "solid-like" atom (S) in that
the particles satisfying this constraint are only allowed to vibrate about
their equilibrium position $\mathbf{r}_{0}$ within their cells$.$ This will be
the situation in the IG, in which the particles do not diffuse out of their
cells. On the other hand, $\overline{\mathbf{r}^{2}}>c_{\text{G}}b$ will
describe a "liquid-like" atom (L) in that the particles escape the
neighborhood of $\mathbf{r}_{0}$ by diffusion and will give rise to a
collective motion. Thus, we can classify the atoms as L or S, thereby reducing
the system to a "fictitious" binary mixture of L and S particles. This
scenario was conjectured by us a while back \cite{GujGld} as the possible
origin of the $\beta$\ relaxation in SCL. This is the only relaxation in IG in
which there are no L-type particles, whereas there are many L-type particles
in SCL contributing to the $\alpha$ relaxation. Hence, we can use the density
$n_{\text{L}}$ of L-type particles as the order parameter to describe the IG
transition: it is non-zero in SCL and gradually vanishes at $T_{\text{K}}$ as
the temperature is reduced; it remains zero in the IG\ state. We can use the
proximity of two L-type particles to determine if they are "connected" to form
a cluster. These clusters can be classified as \emph{liquid-like clusters}
(formed by L particles). One can similarly define a solid-like cluster.
Indeed, IG is a macroscopically large random solid cluster with no liquid-like
clusters embedded in it. At all higher temperatures, both clusters will be
continuously changing in time, but their average densities will remain
constant, similar to what happens in physical gelation.
\cite{Zallen,Flory,deGennes} Therefore, we may be dealing with the phenomenon
of percolation. A very similar picture has been developed by Novikov et al.
\cite{Novikov} Another picture involving two kinds of regions very similar to
the above has been proposed by de Gennes. \cite{deGennesGlass}

Let us now turn to the lattice model. The IG forms a single macroscopic
cluster of solid-like atoms, in which the density $n_{\text{L}}=0$. A real
glass, on the other hand, even at $T_{\text{K}}$, would have a non-zero
$n_{\text{L}}$; the latter would presumably remain non-zero in a real glass
even at $T=0$, and would give rise to an additional energy to the real glass
above $E_{\text{K}}$ and to the residual entropy at absolute zero.
\cite{GujResidualEntropy} Thus, the concept of the order parameter makes sense
only in the context of the ideal glass transition. We, therefore, hope that
our calculation would be able to shed some light on this issue.

\section{Lattice Model\label{Model}}

\subsection{Motivation}

We have already argued in Sect. \ref{Lattice_Model} that the communal entropy
is the central quantity of interest for the glass transition, whose vanishing
locates the glass transition. The configurational contribution $S_{\text{b}%
}(T)$ due to the confinement plays no role in locating the glass transition.
Accordingly, we focus our attention on only on $S_{\text{comm}}(T)$, and as
remarked above in Sect. \ref{Lattice_Model}, we accomplish this easily by
considering a lattice model in which
\[
S_{\text{b}}(T)=0.
\]
The simplest way to obtain the lattice is to replace each cell by a point
called site, and connecting neighboring sites by bonds. The particles or voids
in the system are restricted to be on these sites. The cell-site
transformation results in a lattice structure in which the coordination number
of each site will in general not be the same as shown by the pentagon near the
top right in (a) in Figs. \ref{Fig_Random_Ordered_States0} and
\ref{Fig_Random_Ordered_States}, while it will be the same in the ordered
state as shown in (b) Figs. \ref{Fig_Random_Ordered_States0} and
\ref{Fig_Random_Ordered_States}. If all the cells are occupied by a particle
as shown in Fig. \ref{Fig_Random_Ordered_States0}, then the number of sites is
fixed and equal to $N$, the number of particles. The case when there are empty
cells is shown in Fig. \ref{Fig_Random_Ordered_States}. In this case, the
number of sites will be larger than $N$.

Let us first consider Fig. \ref{Fig_Random_Ordered_States0} for simplicity
with no empty sites. The lowest energy ordered and disordered states are easy
to distinguish because of the disordered lattice in (a). To ensure this
distinction will require considering a disordered lattice to describe a glass,
which creates a complication as it is much easier to deal with a homogeneous
lattice. This is easily taken care of by allowing voids so that the same
homogeneous lattice can describe an ordered and disordered state by simply
having an ordered and disordered distribution of particles and voids. Such a
distinction cannot be made if there were no voids. There exists a potential
energy $E$ for each distribution, and the problem reduces to knowing the
multiplicity, and hence the entropy $S(E)$ of each of the possible energy $E$.

\subsection{Antiferromagnetic Ising Model}

We can use an Ising spin $S$ to represent the particle ($S=+1$) or the void
($S=-1$). One can also use the Ising spin to denote particles A and B of
different species. We take $S=+1$ to denote A, and $S=-1$ to denote B
particles for an alloy or a binary mixture. Thus, the model we use can either
be interpreted as a compressible pure component or an incompressible binary
mixture or an alloy by a slight change in the interpretation. The Ising spins
are located at each site of the lattice. As voids are surrounded by particles
and particles surrounded by voids in the ordered state, this staggered
distribution requires an \emph{antiferromagnetic interaction} between the
Ising spins. We introduce\ the following \emph{AF} \emph{Ising model} in zero
magnetic field on a square or a cubic lattice (lattice spacing $a$) with the
interaction energy%
\begin{equation}
E\mathcal{=}J\sum SS^{\prime}+J^{\prime}\sum SS^{\prime}S^{\prime\prime
},\ \ \ J>0. \label{BinaryE}%
\end{equation}
The first sum is over nearest-neighbor spin pairs and the second over
neighboring spin triplets, which we take to be within a square for simplicity.
The PF is given by%
\begin{equation}
Z(T)=\sum_{\{S=\pm1\}}\exp(-\beta E), \label{PF}%
\end{equation}
where the sum is over all configurations of the $N$ Ising spin states. For
$\left\vert J^{\prime}\right\vert \leq2J$, we have an AF ordering\ at low
temperatures with a sublattice structure:\ spins of a given orientation are
found preferentially on one of the two sublattices. Antiferromagnetically
ordered squares (AFS) with spins alternating, and ferromagnetically ordered
squares (FS) with spins in the same state are the only two square
conformations that contribute to the first term in (\ref{BinaryE}). We may
identify the AF ordered structure as a crystal. \cite{GujS} For two particles
of the same species, this model represents a repulsion at a lattice spacing
$a$. Thus, two particles of the same species prefer to be next-neighbor, with
a particle of a different species at the intermediate site. In CR, such a
situation is preferred. For a glass, three particles of the same species can
occur on neighboring sites, which then raises the energy with respect to CR.
Thus, the model seems to capture the correct physics to give rise to a glass.

For $\left\vert J^{\prime}\right\vert \geq2J,$ the AF ordering is destroyed at
low temperatures; $S$ is the same everywhere at $T=0$. We set $J$=1 to set the
temperature scale and only consider $\left\vert J^{\prime}\right\vert \leq2J$.
It is easy to see that the free energy depends on $\left\vert J^{\prime
}\right\vert ,$ not on\ its sign. In particular, the ground state energy per
spin of the AF ordered state is $E_{0}=-2J,$ regardless of $J^{\prime}.$ In
the following, we will measure the energy and the free energy with respect to
the ground state to give the excitation energies. In this case, both will
vanish at $T=0.$ The non-zero value of $\left\vert J^{\prime}\right\vert
$\ creates a preference for the product $SS^{\prime}S^{\prime\prime}$ in a
square to be of a fixed sign, which then competes with the formation of the
crystal in which this product can be of either sign. A positive $\ ($%
negative$)$ $J^{\prime}$\ provides a preference for $S=-1$ ($+1$), so
$J^{\prime}$ can be used to also control the abundance of one of the spin states.

\subsection{Recursive Lattice}

The entropy $S(T)$ of the model cannot be negative if the state has to occur
in Nature or in simulations; indeed, neither can ever produce a state, which
will exhibit any entropy crisis. If the metastable state entropy $S(T)=0$ at a
positive temperature $T_{\text{K}}$ as the temperature is reduced$,$ then it
must stop there. Its \emph{mathematical extension} to any lower temperature
will necessarily give rise to an entropy crisis and must be replaced by an
\emph{ideal glass} below $T_{\text{K}},$ the ideal glass transition
temperature.\emph{ }\cite{note00} Thus, the partition function for the
metastable state in this case makes physical sense only over $T\geq
T_{\text{K}}$. This is accomplished by restricting the sum in (\ref{PF}) to
disordered configurations foe $E\geq E_{\text{K}}$%
\begin{equation}
Z_{\text{dis}}(T)=\sum_{E\geq E_{\text{K}}}W_{\text{dis}}(E)\exp(-\beta
E),\ W_{\text{dis}}(E_{\text{K}})=1; \label{Dis_PF}%
\end{equation}
see Fig. \ref{Fig-S-T}. The temperature $T_{\text{K}}$ is given by
(\ref{Inverse_Temp}) applied to $S_{\text{dis}}(E)$ at $E_{\text{K}}$. The
above PF is not physically meaningful below $T_{\text{K}}$, even though it can
be mathematically continued to lower temperatures by considering
$W_{\text{dis}}(E)<1$ below $E_{\text{K}}$. As we will see, the method of
calculation reported in Sect. \ref{Binary_Mixture} gives the continuation of
the free energy. This continuation is the origin of a negative entropy that we
report in Sect. \ref{Results}. The mathematical extension describes a stable
state in that the heat capacity remains non-negative. It only suffers from a
negative entropy, something that happens in classical statistical mechanics
such as for ideal gas. This issue has been discussed elsewhere by us.
\cite{GujS}

The model cannot be solved exactly except in one dimension. It is usually
studied in the mean-field approximation commonly known as the Bragg-Williams
approximation \cite{Kubo} adapted for an AF case. However, the approximations
is known to be very crude. Indeed, Netz and Berker \cite{Berker} have shown
that one of the shortcomings of the approximation is that it abandons the
hard-spin condition $S^{2}=1.$ This condition is easily incorporated in exact
calculations on recursive lattices \cite{GujPRL} and it was discovered that
such calculations are more reliable than the conventional mean-field
approximations. Therefore, we adopt the recursive lattice approach here.%
\begin{figure}
[ptb]
\begin{center}
\includegraphics[
trim=0.585214in 0.000000in 0.000000in 0.000000in,
height=3.0009in,
width=3.0268in
]%
{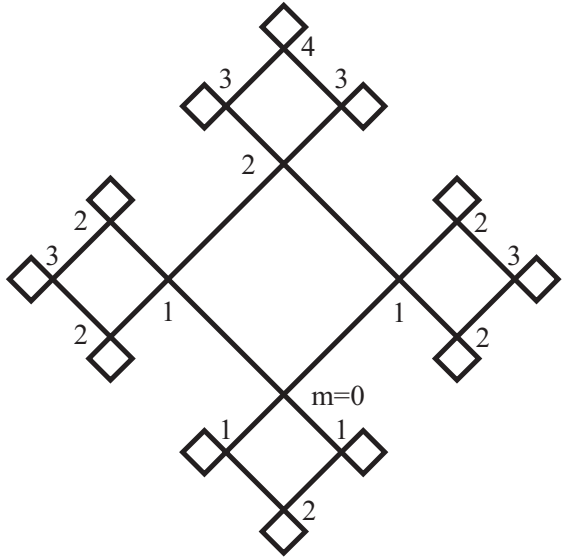}%
\caption{A small portion of a recursive Husimi cactus with two squares ($q=2$)
meeting at each sites. The sites of the squares are labeled as shown, with the
site index increasing as we move away from the origin $m=0.$ }%
\label{Fig_Husimi_Cactus}%
\end{center}
\end{figure}

We consider a Husimi lattice made of squares, on which the model can be solved
exactly. \cite{GujPRL} We consider the simplest lattice shown in Fig.
\ref{Fig_Husimi_Cactus} in which only two squares ($q=2$) meet at a site; they
cannot share a lattice bond. The method is easily extended to consider more
than two squares ($q>2$) meeting at a site. The squares are connected so that
there are no closed loops except those formed by the squares. The lattice can
be thought as an approximation of a square lattice for $q=2$ or a cubic
lattice for $q=3$, so that the exact Husimi lattice solution can be thought of
as an approximate solution of the square lattice model or a cubic lattice,
respectively. Of course, we can also think of the solution as the exact
solution on a recursive lattice, although artificial. The exactness ensures
that stability will always be satisfied. There is a sublattice structure at
low temperatures caused by the anti-ferromagnetic interaction:\ particles of
one species are found on one of the two sublattices. We identify this ordered
structure as a crystal.

\section{Solution\label{Binary_Mixture}}

The method of the solution recursively is standard by now. \cite{GujPRL} We
label sites on the lattice by an index $m$, which increases\ sequentially
outwards from $m=0$ at the origin; see Fig. \ref{Fig_Husimi_Cactus}. We
introduce partial PF's $Z_{m}(\uparrow)$ and $Z_{m}(\downarrow),$ depending on
the states of the spin at the $m$-th lattice level, which represent the
contribution of the part of the lattice above that level to the PF, subject to
the condition that the spin state of the spin at the $m$-th lattice level is
$\uparrow$ or $\downarrow$ respectively. We then construct recursion relations
(RRs) between the partial PFs at level $m$ in terms of the partial PFs at
higher levels $m+1$ and $m+2$ using the standard method.\cite{GujPRL} We
introduce the ratio
\begin{equation}
x_{m}\equiv Z_{m}(\uparrow)/[Z_{m}(\uparrow)+Z_{m}(\downarrow)],
\label{Ratios}%
\end{equation}
which is found to satisfy the recursion relation
\begin{equation}
x_{m}\equiv\frac{f(x_{m+1},x_{m+2},v)}{g(x_{m+1},x_{m+2},v)}, \label{RR}%
\end{equation}
where
\begin{subequations}
\label{Polynomial}%
\begin{align}
f(x,x^{\prime},v)  &  \equiv x^{2r}x^{\prime r}/uv^{2}+2x^{r}x^{\prime r}%
y^{r}v+x^{2r}y^{\prime r}v+ux^{\prime r}y^{2r}+2x^{r}y^{r}y^{\prime r}%
+y^{2r}y^{\prime r}/v,\label{Polynomal_1}\\
g(x,x^{\prime},v)  &  =f(x,x^{\prime},v)+f(y,y^{\prime},1/v),
\label{Polynomal_0}%
\end{align}
with $r=q-1$\ and where%
\end{subequations}
\[
u\equiv e^{4\beta},v\equiv e^{2\beta J^{\prime}},y\equiv1-x,y^{\prime}%
\equiv1-x^{\prime}.
\]

There are two kinds of fix-point (FP) solutions of the recursion relation
(\ref{RR}) that describe the bulk behavior. \cite{GujPRL} In the 1-cycle
solution, the FP solution becomes independent of the level $m$, and is
represented by $x^{\ast}.$ It is given by%
\[
x^{\ast}\equiv\frac{f(x^{\ast},x^{\ast},v)}{g(x^{\ast},x^{\ast},v)}%
\]
For $J^{\prime}=0$, $x^{\ast}$ is given by $x^{\ast}=1/2$, as can be checked
explicitly. For $J^{\prime}\neq0,$ $x^{\ast}\neq1/2$ and has to be obtained
numerically. This solution \emph{exists} at all temperatures $T\geq0;$ thus,
there is no spinodal of this solution$.$ This solution describes the
disordered phase. The other FP solution of interest is a 2-cycle solution
associated with the AF state containing AFSs. \cite{GujPRL} It alternates
between two values $x_{1}^{\ast},$ and $x_{2}^{\ast}$ which occur at
successive levels. It is given by%
\[
x_{1}^{\ast}\equiv\frac{f(x_{2}^{\ast},x_{1}^{\ast},v)}{g(x_{1}^{\ast}%
,x_{2}^{\ast},v)},\ x_{2}^{\ast}\equiv\frac{f(x_{1}^{\ast},x_{2}^{\ast}%
,v)}{g(x_{2}^{\ast},x_{1}^{\ast},v)}.
\]
This kind of FP solution has also been observed in other systems such as
semi-flexible polymers \cite{GujC,GujRC}, dimers, \cite{GujS} and stars and
dendrimers, \cite{CorsiThesis} and has been thoroughly investigated. At $T=0,$
the 2-cycle solution is given by $x_{1}^{\ast},x_{2}^{\ast}=1,0$ or $0,1$
describing the perfect crystal. This solution then evolves with $T$ due to
excitations and describes the crystal at low temperatures. The free energy is
calculated by the general method due to Gujrati.
\cite{GujPRL,GujC,CorsiThesis} Whichever solution has the lower free energy
represents the equilibrium state. The solution with the higher free energy,
then, represents SMS. Both solutions can only be observed in Nature if their
entropy remains non-negative. The temperature where the two solutions have the
same free energy is the transition temperature, which we denote by
$T_{\text{M}}$.%
\begin{figure}
[ptb]
\begin{center}
\includegraphics[
trim=0.859026in 3.726438in 1.436809in 2.870379in,
height=3.1038in,
width=4.3682in
]%
{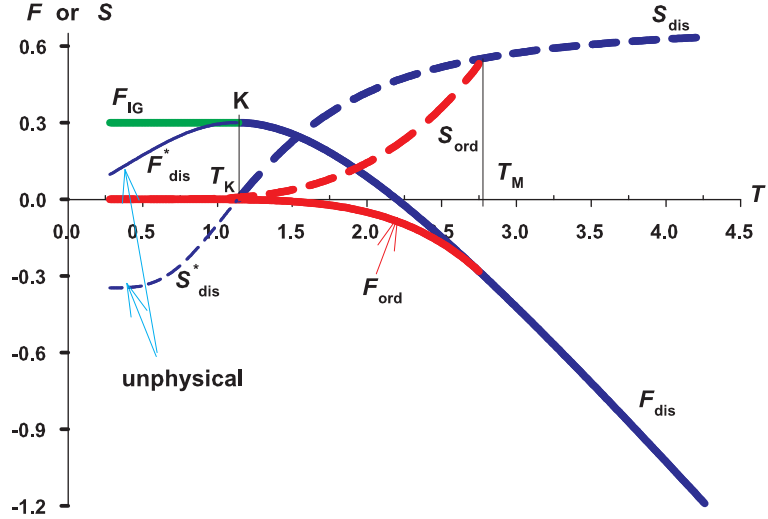}%
\caption{The free energy and entropy for the two FP solutions. The model shows
an entropy crisis and an ideal glass transition at $T_{\text{K}}.$ The thin
curves, indicated by an asterisk, represent unphysical states below
$T_{\text{K}}$ (negative entropy) and are replaced by the ideal glass state
IG.}%
\label{F1}%
\end{center}
\end{figure}

Let $\phi_{\text{FS}}$ and $\phi_{\text{AFS}}$ denote the density of squares
per site in which the four spins are ferromagnetically ordered ($\uparrow
\uparrow\uparrow\uparrow$ or $\downarrow\downarrow\downarrow\downarrow$) and
antiferromagnetically ordered ($\uparrow\downarrow\uparrow\downarrow$),
respectively. These densities are calculated below; see
(\ref{FS-AFS_densities}). At $T\rightarrow\infty,$ all spins are uncorrelated
so that the density per site $\phi_{\text{FS}}=1/16,$ and $\phi_{\text{AFS}%
}=1/16,$ and the entropy per spin is $S=\ln2.$ As $T$ is reduced,
$\phi_{\text{FS}}$ decreases, while $\phi_{\text{AFS}}$ increases. For the
perfect crystal, which occurs at absolute zero, $\phi_{\text{FS}}=0$%
,$\phi_{\text{AFS}}=1/2$.

\section{Results\label{Results}}

\subsection{Ideal Glass transition}

\begin{figure}
[ptb]
\begin{center}
\includegraphics[
trim=0.000000in 1.433131in 0.860722in 0.575334in,
height=3.4791in,
width=4.3647in
]%
{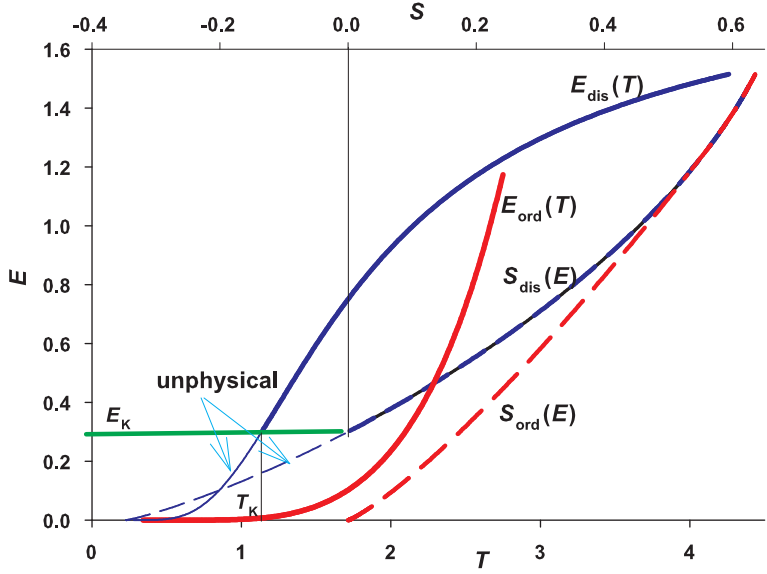}%
\caption{$S-E-T$ relationship for the two FP solutions. The excitations in the
two solutions near $T=0$ are very different. The excitations in the 1-cycle
state near $T_{\text{K}}$ are strongly interacting and correlated in the form
of FSs as opposed to those near $T=0.$}%
\label{F2}%
\end{center}
\end{figure}

The results for $q=2$ are presented in Figs. \ref{F1}-\ref{F3}, where all the
quantities are defined as per spin (or particle). The free energy
$F_{\text{dis}}$ and entropy $S_{\text{dis}}$ (per spin) associated with the
1-cycle FP solution are shown by the continuous and the long dash blue curves
in Fig. \ref{F1}. The free energy $F_{\text{ord}}$ and entropy $S_{\text{ord}%
}$ associate with the 2-cycle FP solution are shown by the continuous and the
long dash red curves. The energy $E(T)$ as a function of $T$ and the entropy
$S(E)$ as a function of $E$ are shown in Fig. \ref{F2}. We have set
$J^{\prime}=0.01$ for Figs. \ref{F1} and \ref{F2}. The energy $E$ and the free
energy $F$ are defined so that they represent the contributions of excitations
with respect to the ground state energy $E_{0}=-2J=-2$, so that they vanish at
$T=0,$ as is clearly seen in Figs. \ref{F1}, and \ref{F2}. The transition
temperature from disordered state to the ordered state is found to be
$T_{\text{M}}\cong2.7706$. We see from Fig. \ref{F1} that $F_{\text{dis }}%
$crosses zero and becomes positive below $T=T_{\text{eq}}\simeq2.200$. This is
in accordance with Fig. \ref{Fig-S-T}; see line OO$^{\prime}$. The free energy
$F_{\text{dis }}$\ again becomes zero (not shown here, but we have checked it)
as $T\rightarrow0;$ compare it with the free energy in the inset in Fig.
\ref{Fig-S-T}. Thus, $F_{\text{dis }}$ possesses a maximum at an intermediate
temperature (see point K in Fig. \ref{F1}) at $T=T_{\text{K}}\simeq1.1316,$ so
that the entropy $S_{\text{dis}}$ vanishes there. Another interesting
observation is that $F_{\text{dis}}(T=0)=F_{\text{ord}}(T=0)=0$. We have seen
this to hold in all our previous calculations
also.\cite{GujC,GujRC,Fedor,CorsiThesis,GujS,Keith,Cerena}

The calculation for the disordered metastable state in our recursive method
also yields the disordered quantities below $T_{\text{K}}$, even though the PF
in (\ref{Dis_PF}) is not physically meaningful because of the negative
entropy. Below $T_{\text{K}}$, the mathematical \emph{functions}
$F_{\text{dis}}$ and $S_{\text{dis}}$, shown by their thin portions in Figs.
\ref{F1} and \ref{F2}, in our calculation continue to satisfy the stability
condition (non-negative specific heat). Despite this, they do not represent
any physical states in the system due to negative entropy and have to be
discarded as \emph{unphysical}. Below $T_{\text{K}}$, we must
connect\cite{GujS} the metastable state (described by $F_{\text{dis }}$ and
$S_{\text{dis}}$ between $T_{\text{K}}$ and $T_{\text{M}})$ by a glassy phase
of a constant free energy $F=F_{\text{IG}},$ see the green horizontal line in
Fig. \ref{F1}, and $S_{\text{dis}}=S_{\text{IG}}=0.$ The 1-cycle energy at K
is $E_{1\text{K}}=F_{\text{IG}}\simeq0.301.$ The entropy $S_{\text{ord}}$ is
never negative, and the 2-cycle FP solution represents the equilibrium crystal
below $T_{\text{M}}.$ We should observe that $S_{\text{ord}}\geq S_{\text{IG}%
}=0$ over $T<T_{\text{K}}\simeq1.1316$. The same also holds for the Flory
model of semiflexible polymers studied by our group, \cite{GujC,GujRC} and for
anisotropic dimer model. \cite{Fedor,GujS} However, it does not happen in the
calculation by Gibbs and DiMarzio \cite{GibbsDiMarzio}\ because of the poor
approximation. The specific heat of the ordered phase is much higher than that
of the disordered phase, so that the entropy of CR at low temperatures near
$T_{\text{K}}$ is very small. We also observe that the behavior of the entropy
in the ordered and disordered states in Fig. \ref{F2} is consistent with that
in Fig. \ref{Fig-S-T}. In particular, $S_{\text{dis}}$ has a positive slope at
$E_{\text{K}}$ in accordance with Fig. \ref{Fig-S-T}. This feature is
responsible for the entropy crisis at a positive temperature $T_{\text{K}}$.
The same is true of the behavior of the free energy in Fig. \ref{F1}, which is
consistent with that reported in the inset in Fig. \ref{Fig-S-T}. The free
energy has a maximum at $T_{\text{K}}>0$.

\subsection{\textbf{Competition with Crystal Ordering }}

\begin{figure}
[ptb]
\begin{center}
\includegraphics[
trim=1.023864in 3.497862in 1.471646in 3.023129in,
height=3.0554in,
width=4.0568in
]%
{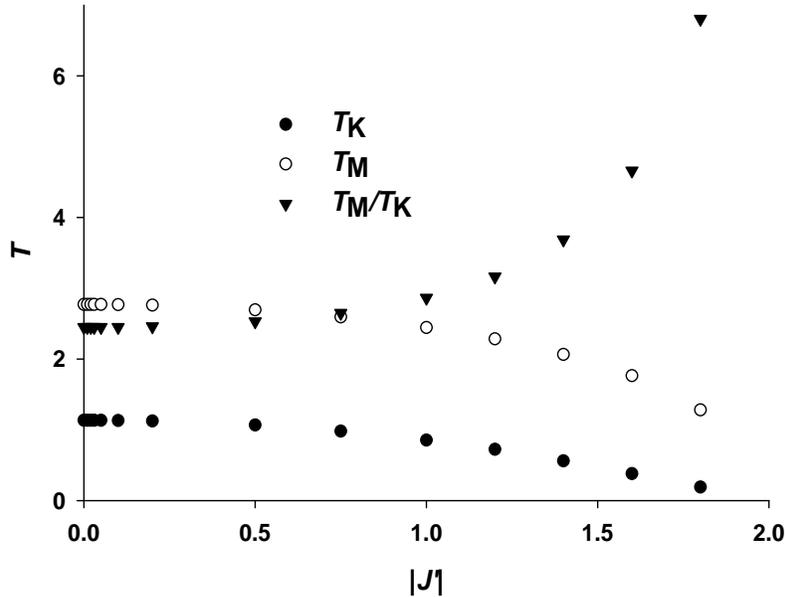}%
\caption{The effect of $\left\vert J^{\prime}\right\vert ,$ which creates
competition with the crystal ordering, on $T_{\text{M}},$ and $T_{\text{K}}$
and their ratio. The weakening of crystal ordering also "weakens" the ideal
glass formation. }%
\label{F3}%
\end{center}
\end{figure}

The behavior as a function of $\left\vert J^{\prime}\right\vert $ of the
transition temperature $T_{\text{M}}$ (empty circles), the ideal glass
transition temperature $T_{\text{K}}$ (filled circles) and their ratio
$T_{\text{M}}$/$T_{\text{K}}$ (triangles) are shown in Fig. \ref{F3}. As said
above, $\left\vert J^{\prime}\right\vert $ competes with the crystal ordering
and reduces $T_{\text{M}}.$ One can use the inverse ratio $T_{\text{K}}%
$/$T_{\text{M}}$ as a \emph{measure} of the relative ease of glass formation:
larger this value, easier it is to obtain the ideal glass as $T_{\text{K}}$ is
not too deep relative to $T_{\text{M}}.$ This relative depth is measured by
the ratio $T_{\text{K}}$/$T_{\text{M}}$. What we observe is that $T_{\text{M}%
}$/$T_{\text{K}}$ increases with $\left\vert J^{\prime}\right\vert ,$ with
$T_{\text{K}}$ approaching zero faster than $T_{\text{M}},$ so that the ratio
$T_{\text{M}}$/$T_{\text{K}}$ continues to increase with $\left\vert
J^{\prime}\right\vert .$ This implies that it becomes harder to obtain the
ideal glass as $T_{\text{K}}$ becomes relatively farther away from
$T_{\text{M}}$ as $\left\vert J^{\prime}\right\vert $ increases$.$ The
competition provided by $\left\vert J^{\prime}\right\vert $ weakens not only
crystal ordering but also "weakens" forming the ideal glass. Consequently,
competition does not enhance the ability to undergo ideal glass transition, an
interesting result which is being explored further \cite{Cerena} to see if
other competitions behave similarly.

We have considered a small value of $J^{\prime}$ in order to compare the
results with the simple case $J^{\prime}=0$, which is analytically solvable;
see Sect. \ref{Defects}. The complete analysis for other values of $J^{\prime
}$ will be reported later. \cite{Keith} Here, we quote the results for a
larger value of $J^{\prime}=1.6$. The melting temperature $T_{\text{M}}%
\cong1.762$, see Fig. \ref{F3}, where $S_{\text{dis}}\cong0.40999$ and
$S_{\text{ord}}\cong0.40947,$ and the shifted free energy is $F\cong$
$-0.299026$. The ideal glass occurs at $T_{\text{K}}\cong0.377,$ as shown in
Fig. \ref{F3}. We notice that the entropy discontinuity is not very large at
$T_{\text{M}}$; however, the entropy of the ordered phase drops very fast just
below $T_{\text{M}}$ so that around $T_{\text{K}}$, the CR entropy is quite
small (result not shown); compare with Fig. \ref{F1}, where the drop is
similarly fast.%

\subsection{\textbf{Analysis of Defects\label{Defects}}}

\subsubsection{Disordered Phase}

To understand the difference between the defects in the disordered liquid and
the crystal, we turn to Fig. \ref{F2} and observe that near $T=0,$ the
excitation energies of both FP solutions are very different, even though
$E_{\text{dis}}(0)=E_{\text{ord}}(0)=0$. Detailed analysis will be presented
elsewhere. \cite{Keith} The excitations (defects) in the crystal are known to
be due to \emph{point-like} excitations caused by the reversal of a single
spin which changes the free energy by $\simeq8J$ (coordination number 4) with
respect to the ground state$;$ here we assume $J^{\prime}$ to be small.
Therefore, this excitation causes the leading term\cite{Domb} in the free
energy $F_{\text{ord}}$ to be $1/u^{2}$ and can be treated as
\emph{non-interacting} as long as they are small in number. The exact
recursive method also allows us to calculate the densities of various
excitations and defects. \cite{Keith} In particular, the densities
$\phi_{\text{FS}}$ and $\phi_{\text{AFS}}$ are easy to calculate. For the
$2$-cycle solution ($x_{1}^{\ast},x_{2}^{\ast},x_{1}^{\ast},x_{2}^{\ast
},\cdots$), they are given by%
\begin{equation}
\phi_{\text{FS}}=\frac{x_{1}^{\ast2r}x_{2}^{\ast2r}/v^{2}+y_{1}^{\ast2r}%
y_{2}^{\ast2r}v^{2}}{2ug(x_{1}^{\ast},x_{2}^{\ast},v)(x_{\alpha}^{\ast
q}+y_{\alpha}^{\ast q})},\ \phi_{\text{AFS}}=\frac{[x_{1}^{\ast2r}y_{2}%
^{\ast2r}+y_{1}^{\ast2r}x_{2}^{\ast2r}]u}{2g(x_{1}^{\ast},x_{2}^{\ast
},v)(x_{\alpha}^{\ast q}+y_{\alpha}^{\ast q})},\ \ \alpha=1,2.
\label{FS-AFS_densities}%
\end{equation}
The densities per site of up ($\uparrow$) and down ($\downarrow$) spins are
given by%
\begin{equation}
\phi_{\text{up}}=\frac{x_{\alpha}^{\ast q}}{x_{\alpha}^{\ast q}+y_{\alpha
}^{\ast q}},\ \phi_{\text{down}}=\frac{y_{\alpha}^{\ast q}}{x_{\alpha}^{\ast
q}+y_{\alpha}^{\ast q}},\ \ \alpha=1,2. \label{up-down_densities}%
\end{equation}
By setting $x_{1}^{\ast}=x_{2}^{\ast}=x^{\ast},$ the same expressions will
also give the values of $\phi_{\text{FS}}$ and $\phi_{\text{AFS}}$\ or
$\phi_{\text{up}}$ and $\phi_{\text{down}}$ for the disordered state.

What kinds of excitations are deducible from the form of $F_{\text{dis}}$ near
$T=T_{\text{K}}?$ To answer this, we consider the simple case of $J^{\prime
}=0.$ In this case, the $1$-cycle solution is given by $x^{\ast}=1/2$ at all
temperatures. This solution also gives us $F_{\text{dis}}$ over the entire
temperature range (physical and unphysical) $T\geq0.$ As we are only
interested in the excitation energy, we have shifted the free energy so that
it vanishes at $T=0.$ The unphysical nature of this free energy (negative
entropy) between $T=0$ and $T=T_{\text{K}}$ does not affect the energy as the
specific heat is non-negative; it only affects the entropy. This portion of
the free energy will correctly give the information about the energy and the
nature of excitations at $T_{\text{K}}$ or above$.$ From now on, we consider
$r=1$. The shifted free energy is given by%
\[
F_{\text{dis}}(T)=-(T/2)\ln w+2,
\]
where we have introduced%
\[
w=3+u/2+1/2u=4g(1/2,1/2,1),
\]
where $g(1/2,1/2,1)$ is the value of $g(x,x^{\prime},v)$\ in
(\ref{Polynomal_0}) at the $1$-cycle fix-point for $v=1$. The excitation
energy is obtained from the free energy, and is given by%
\[
E_{\text{dis}}(T)=-(u-1/u)/w+2,
\]
which can be written as
\[
E_{\text{dis}}(T)=4(\phi_{\text{FS}}-\phi_{\text{AFS}})+2,
\]
in terms of $\phi_{\text{FS}}$ and $\phi_{\text{AFS}}$, which from
(\ref{FS-AFS_densities}) are given by
\[
\phi_{\text{FS}}=1/4uw,\ \ \phi_{\text{AFS}}=u/4w
\]
\ for the $1$-cycle solution. The entropy is calculated using
\[
S_{\text{dis}}(T)=(E_{\text{dis}}-F_{\text{dis}})/T.
\]
The densities of up and down spins are equal at all temperatures%
\[
\phi_{\text{up}}=\ \phi_{\text{down}}=1/2.
\]
Thus, the disordered state corresponds equal number of up and down spins at
all $T\geq0$. At $T=0,E_{\text{dis}}(0)=0,$ which implies that all squares are
AFS. Thus, as the temperature is raised, more and more FS's appear by changing
a AFS into a FS. Near $T=0,$ we find that
\begin{align*}
E_{\text{dis}}(T)  &  \simeq12/u,~\\
F_{\text{dis}}(T)  &  \simeq T(%
\frac12
\ln2-3/u),\\
S_{\text{dis}}(T)  &  \simeq-%
\frac12
\ln2+3/u+12\beta/u.
\end{align*}
At $T=0$, $S_{\text{dis}}(0)$ in Fig. \ref{F1} is found to be almost $-%
\frac12
\ln2$. What we discover is that the excitations due to $1/u$-term near $T=0$
are not the \emph{uncorrelated} single spin reversal in the background of a
perfect crystal; the latter require an excitation energy of $8J$ and a
correction of order $1/u^{2}$ in the free energy of the ordered state. Rather,
they represent \emph{correlated} excitations in the form of FS in the
background of AFSs by turning an AFS into a FS. Each FS excitation requires an
energy $4J$ per site. This is in accordance with the general discussion above.
Indeed, the excitation spectrum is given by the expansion \cite{Domb} of
$E_{\text{dis}}(T)$ in powers of $1/u$. These excitations also explain why the
thin portion of $E_{\text{dis}}(T)$ rises more rapidly than $E_{\text{ord}%
}(T)$ in Fig. \ref{F2} so that $E_{\text{K}}\simeq0.3$ is appreciably higher
than the $E_{\text{ord}}(T)\simeq0.1$ at $T_{\text{K}}$ due to a lower
$\phi_{\text{AFS}}$ and higher $\phi_{\text{FS}}\simeq3.58\times10^{-4}$ in
the metastable state%
\[
\phi_{\text{AFS}}\simeq0.426,\ \phi_{\text{FS}}\simeq3.58\times10^{-4},
\]
relative to
\[
\phi_{\text{AFS}}\simeq0.498,\ \phi_{\text{FS}}\simeq7.991\times10^{-7}%
\]
in the crystal. The density of sites
\[
\phi_{\text{unc}}\equiv1-2(\phi_{\text{FS}}+\phi_{\text{AFS}})
\]
not covered by AFSs and FS's is
\begin{align*}
\phi_{\text{unc}}  &  \simeq0.148\text{ in SCL,}\\
\phi_{\text{unc}}  &  \simeq3.684\times10^{-3}\text{ in CR.}%
\end{align*}
These sites are probably uncorrelated in CR to account for non-zero entropy of
CR at $T_{\text{K}}$, but this needs to be carefully checked. However, these
sites and the two kinds of squares must be strongly correlated in SCL and IG
at $T_{\text{K}}$ to ensures that the entropy is zero.

\subsubsection{Ordered Phase}

The CR entropy $S_{\text{ord}}(T)$ vanishes only at absolute zero, but remains
non-zero at any other temperature. This is consistent with defects in the
crystal near $T=0$ being point defects. As the lattice entropy represents the
communal entropy, it is not zero for CR. As a consequence, CR is \emph{not}
confined to a single basin, as required by (\ref{S_comm_CR}), due to the
presence of point defects that emerge in the ideal CR. The presence of point
defects gives rise to basins different from the basin corresponding to the
perfect CR at absolute zero. In a real CR, there will be additional
contribution $S_{\text{b}}(T)$ due to vibrations inside these basins. The
point defects form a dilute system near $T=0$, so that there is practically no
correlation among them. Thus, as discussed earlier in Sect. \ref{Non-zero TK},
the derivative in (\ref{Inverse_Temp}) is infinitely large for the perfect CR.

\subsection{Nature of the Ideal Glass}

As there is no non-analyticity in the computed (physical and unphysical)
$F_{\text{dis}}(T)$\ of SCL at $T_{\text{K}}$, the excitation spectrum remains
continuous at $T_{\text{K}}$. Thus, it can be continued above $T_{\text{K}}$
where physical states (non-negative entropies) occur. The above analysis leads
us to conclude that the excitations at or above $T_{\text{K}}$ in SCL are very
different from the point defects of the crystal and are the ones that get
frozen in the ideal glass that is formed at $T_{\text{K}}.$ For $T<T_{\text{K}%
},$ we have an ideal glass, shown by the horizontal short dash curve in Fig.
\ref{F2}, of \ constant energy $E_{\text{K}}$ and zero entropy. Let us try to
understand how the entropy of a disordered state can be zero. Let us consider
a system consisting of $N$ Ising spins. Let $W_{\text{dis}}$ denote the number
of disordered microstates, which we index by $i=1,2,\cdots,W_{\text{dis}}$.
Let $E_{i}$ denote the energy (and not the energy per site) of the $i$th
microstate, and $p_{i}$\ its probability to occur. The latter is given by
\[
p_{i}=e^{\beta(NF_{\text{dis}}-E_{i})},\ (%
{\textstyle\sum}
p_{i}=1),
\]
and the entropy per spin $S_{\text{dis}}$ is given by
\[
NS_{\text{dis}}=-\sum_{i=1,W_{\text{dis}}}p_{i}\ln p_{i}.
\]
The entropy can be zero \cite{note0} if and only if one of the probabilities
is unity, while all other probabilities are zero. Let $i=i_{0}$ denote the
particular disordered microstate for which
\[
p_{i_{0}}=1.
\]
This particular disordered microstate, being the only one of energy $E_{i_{0}%
}$, also represents the \emph{macrostate} identified as the ideal glass at
$T\leq T_{\text{K}}.$ The energy of this microstate per spin is%
\[
E_{i_{0}}/N\equiv E_{\text{K}}\equiv F_{\text{IG}}.
\]
This discussion clearly shows how the entropy of IG can be less than that of
the CR at the same temperature $T\leq T_{\text{K}}$. There is nothing
paradoxical here. The analysis also shows that the microstate $i_{0}$ emerges
out of the above defects frozen at $T_{\text{K}}$. These defects also
determine the nature of solid-like clusters discussed in Sect.
\ref{Order_Parameter} that survive at $T_{\text{K}}.$ These clusters pack the
lattice in a \emph{unique} way to generate the microstate $i_{0}$, so that the
resulting entropy vanishes. From what has been said above, these clusters are
strongly correlated. How this correlation comes about as SCL approaches
$T_{\text{K}}$ is something we are not able to answer at present and is under
investigation. Understanding the nature of this unique microstate is certainly
a challenging problem in theoretical physics at present.

\subsection{Excitation Spectra in Real Glasses}

The excitation spectra of both solutions over the physical range are
completely described by their respective entropies $S_{\text{dis}}(E)$ (the
thick part) and $S_{\text{ord}}(E)$ shown in Fig. \ref{F2}. For $E<E_{\text{K}%
},$ the excitations in IG cannot change since they are frozen at constant
energy $E_{\text{K}}$ into the unique microstate $i_{0},$ but continue to
change in the crystal. In experiments, the ideal glass will never be observed
due to time-limitations and one would obtain a non-stationary state, see the
green curve GF in Fig. \ref{Fig-S-T}, whose entropy $S_{\text{NEQS}}(E)$ must
satisfy $S_{\text{NEQS}}(E)\leq$ $S_{\text{dis}}(E)$ according to the law of
increase of entropy. \cite{Landau} It can be shown \cite{Gujrati-Relaxation}
that at a given temperature, the energy and entropy of the a NEQS continue to
decrease in time as the state approaches SMS. Thus, the non-stationary glass
will have some extra excitations at low temperatures with respect to the IG at
that temperature. These defects are the one that we have associated with
$n_{\text{L}}$ in Sect. \ref{Order_Parameter}. The excitations just above
$T_{\text{K}}$ as IG "melts" into SCL are also due to the liquid-like clusters
introduced in Sect. \ref{Order_Parameter}, which must be strongly correlated.
The associated excitation energy of these clusters is given by the behavior of
$E_{\text{dis}}(T)$ above $T_{\text{K}}.$

The down spins (voids) distribute themselves in the lattice at equilibrium,
and the corresponding 1-cycle solution gives rise to an excitation spectrum so
that $S_{\text{dis}}(E)$ vanishes at $T_{\text{K}}.$ If it happens that the
system is quenched, then all we can say is that the corresponding spectrum
$\widetilde{S}_{\text{dis}}(E)$ of the quenched system, which should be
similar to the green curve GF in Fig. \ref{Fig-S-T}, must satisfy the standard
condition $\widetilde{S}_{1}(E)\leq$ $S_{1}(E)$. Despite this, the entropy of
the quenched system does not vanish at a positive temperature
\cite{Gujrati-Relaxation}. There is no contradiction.

\subsection{Ferromagnetic Case}

For the AF case that we consider here, the 1-cycle solution is found to exist
at \emph{all} temperatures and describes the disordered liquid above and its
metastable continuation below the transition temperature. There is no
singularity in this fix-point solution at the transition. In contrast, for the
ferromagnetic case ($J<0$), the 2-cycle FP solution is never stable, and the
1-cycle solution has a singularity at the ferromagnetic transition and its
entropy never becomes negative. This is not surprising as the ferromagnetic
Ising model is suitable to describe liquid-gas transition which is known to
have a critical point. Thus, this model is not suitable to study supercooling.

\section{Summary}

To summarize, our model calculation demonstrates that monatomic systems also
give rise to an ideal glass and an energy gap. This was one of our two goals
and, in conjunction with the earlier work of Gibbs and DiMarzio
\cite{GibbsDiMarzio} and from our group \cite{GujC,GujRC} for long polymers
and for anisotropic dimer model,\cite{Fedor,GujS} strongly suggests that the
entropy crisis is ubiquitous in molecules of all sizes. The other goal was to
learn about generic properties of ideal glasses such as the nature of defects.
To this end, we have verified that our proposed energy gap model in Sect.
\ref{Singularity_Assumption}, which is consistent with our previous exact
calculations\cite{GujGold,GujC,GujRC,Fedor,GujS,GujMeta,Gujun} is also
consistent with our current model calculation of monatomic glass formers. The
macrostate associated with the ideal glass contains only one microstate
because its entropy is zero. This fact makes defects in IG to be highly
correlated so that the energy of excitation and the resulting entropy are
proportional to each other; see (\ref{Entropy_Gain}) and (\ref{IG_Excitation}%
). The constant of proportionality is $T_{\text{K}}$. Thus, the strongly
correlated defects in glass at $T_{\text{K}}$ are very different from those in
the crystal, in which they are mostly uncorrelated neat $T=0$. Because of the
strong correlation, the defects in IG cannot be removed to lower its energy
and still leave the microstate disordered. Of course, one can remove a lot of
defects to bring its energy to that of CR at any themperature below
$T_{\text{K}}$, but this will convert IG to an ordered state. The enrgy
required to do so will be extensive and its probability will be almost zero at
such temperatures. 

The defects in real glasses will be local, as discussed in Sect.
\ref{Non-zero TK}. Accordingly, one can have a real glass approach absolute
zero, where its entropy need not vanish. As the glass relaxes, its entropy and
free energy will continue to decrease as shown elsewhere.
\cite{Gujrati-Relaxation} Eventually, its entropy will vanish. During
relaxation, the size of the defects will continuously grow until they become
non-local. The time required to completely relax is probably so long compared
to experimental time scales that a real glass will never completely relax.

The existence of the gap is used to prove that IG in SCL that emerges in any
generic system must appear at a positive and finite temperature $T_{\text{K}}%
$, which is seen in all exact recursive lattice calculations
\cite{GujC,GujRC,Fedor,GujS,GujMeta,Gujun} from our group, and the exact
calculation in a one-dimensional polymeric system.\cite{GujMeta,Gujun} The
proof presented in this work is more direct and simple compared to an earlier
proof, which was somewhat involved.\cite{Gujun}

We use the relative depth of $T_{\text{K}}$ below the melting temperature
$T_{\text{M}}$ as a measure of ease of forming a glass. A smaller value of
this ratio is taken to imply that it is more easy to form a glass, as one does
not have to supercool too much below the melting temperature $T_{\text{M}}$.
The ratio is related to the relative depth $(T_{\text{M}}-T_{\text{K}%
})/T_{\text{M}}$. Our model calculation shows that the competition does not
necessarily enhance the ability to form a glass.

We have been unable to completely specify the actual IG microstate, except to
note that its energy cannot be reduced by removing some of its defects and
that its defects cannot be interchanged because that will generate another
microstate, different from IG. The latter will violate the uniqueness of IG
(zero entropy). However, it would be desirable to obtain more specific
information about this unique microstate. It remains a challenge in
theoretical physics to even decide how such a disordered microstate should be
characterized. The work we have reported uses a classical model. It would be
interesting to see if a similar calculation can be carried out for a quantum
system in which configurational degrees of freedom cannot be separated from
the kinetic degrees of freedom. Such a separation was required for the
introduction of the communal entropy. A quantum model calculation will clarify
if the communal entropy can still be defined and if it has any relevance for IG.


\bigskip

\begin{thebibliography}{99}                                                                                               %
\bibitem {Kauzmann}W. Kauzmann, Chem. Rev., \textbf{43}, 219-256 (1948).

\bibitem {Landau}L.D. Landau and E.M. Lifshitz, Statistical Physics,
3$^{\text{rd}}$ edition, Part I, Pergamon Press, Oxford (1986).

\bibitem {Note-Continuation}For example, the singularity in the equilibrium
free energy at the melting transition may not occur when we extend the liquid
state into its supercooled state by restricting allowed microstates to be
disordered. The model is used to describe liquid gas transition. \cite{Note01}
This should be contrasted with how an essential singularity appears in the
droplet model, \cite{Fisher,Suto} where one does not restrict the microstates
in the partition function. The presence of such a singularity is customarily
used to argue that no continuation is possible to the other side of the
singularity. This is not necessarily correct. Limitations of the droplet model
are discussed by Domb. \cite{Domb0} An example of a function with an essential
singularity, but which exists on both sides of the singularity, is the
following real function of a real variable $x$%
\[
f_{A}(x)=Ax^{2n}\exp(-1/x^{2}),\ A,n\text{ are constants,}%
\]
which is singular at $x=0$, but exists on both sides of the singularity. Of
course, one can patch two functions at $x=0$ with different $A$ for either
sides of $x=0$ to construct an infinite number of functions, but this is
irrelevant for the existence of $f_{A}(x).$ In principle, any thermodynamic
function is uniquely obtained by taking the thermodynamic limit. For example,
the above function can emerge from
\[
f_{A,N}(x)=Ax^{2n}\exp[-1/(x^{2}+\alpha/N^{m})],
\]
as $N\rightarrow\infty$ ($m>0;\alpha$ a constant), and is \emph{uniquely}
defined on both sides of $x=0$.

\bibitem {Fisher}M.E. Fisher, Physics, \textbf{3}, 255 (1967).

\bibitem {Suto}A. S\"{u}t\~{o}, J. Phys. A \textbf{15}, L749 (1982).

\bibitem {Note01}The two coexisting phases in liquid-gas transitions have
identical symmetry in that they are simply related by a symmetry operation
like the up-down symmetry in the Ising model. No symmetry operation can
transform crystals into liquids, and vice versa.

\bibitem {Domb0}C. Domb, \textit{Critical Point}, Taylor and Francis, London, (1996).

\bibitem {Bragg}W.L. Bragg and E.J. Williams, Proc. Roy. Soc. \textbf{A145},
699 (1934).

\bibitem {GujMeta}P.D. Gujrati, cond-mat/0412757.

\bibitem {Gujun}P.D. Gujrati, cond-mat/0309143; cond-mat/0404748.\ 

\bibitem {Gujrati-Relaxation}P.D. Gujrati, arXiv:0910.0026.\ 

\bibitem {Maxwell}J.C. Maxwell, Scientific Papers, ed. W.D. Niven (Dover, N.Y.
1965); p.425.\ 

\bibitem {GujGld}P.D. Gujrati and M. Goldstein, J. Phys. Chem. \textbf{84},
859 (1980), and references there in.

\bibitem {note00}Even though no experiment can ever detect the ideal glass
transition due to time constraints, its existence most certainly implies that
real glasses would be formed during experiments.

\bibitem {GibbsDiMarzio}J. H. Gibbs and E. A. DiMarzio, J. Chem. Phys.
\textbf{28}, 373 (1958).

\bibitem {GujGold}P.D. Gujrati, J. Phys. A \textbf{13}, L437 (1980); P.D.
Gujrati and M. Goldstein, J. Chem. Phys. \textbf{74}, 2596 (1981); P.D.
Gujrati, J. Stat. Phys. \textbf{28}, 241 (1982).

\bibitem {GujC}P.D. Gujrati and A. Corsi, Phys. Rev. Lett.\textbf{87}, 025701
(2001); A. Corsi and P.D. Gujrati, Phys. Rev. E \textbf{68}, 031502 (2003); cond-matt/0308555.

\bibitem {GujRC}P. D. Gujrati, S. S. Rane and A. Corsi, Phys. Rev E
\textbf{67}, 052501(2003).

\bibitem {GujS}F. Semerianov and P.D. Gujrati, Phys. Rev. E\textbf{ 72},011102
(2005); cond-mat/0410488; cond-mat/0401047.

\bibitem {Fedor}F. Semerianov, Ph.D. Dissertation, University of\ Akron (2004).

\bibitem {Morgenstern}I. Morgenstern, IBM J. Res. Develop. \textbf{33}, 307 (1989).

\bibitem {Ramirez}A. P. Ramirez, Annu. Rev. Maer. Sci. \textbf{24}, 453 (1994).

\bibitem {SimhaGoldstein}\textit{The Glass Transition and the Nature of the
Glassy State,} edited by M. Goldstein and R. Simha, Ann. N. Y. Acad. Sci.
\textbf{279} (1976).

\bibitem {Goldstein-ExcessEntropy}M. Goldstein, J. Chem. Phys. \textbf{64},
4767 (1976), and references there in.

\bibitem {Goldstein}M. Goldstein, J. Chem. Phys. \textbf{51}, 3728 (1969).

\bibitem {Lennard-JonesDevonshire}J.E. Lennard-Jones and A.F. Devonshire,
Proc. Roy. Soc. (London) \textbf{A163}, 53 (1937).

\bibitem {Kirkwood}J.G. Kirkwood, J. Chem. Phys. \textbf{18}, 380 (1950); P.
Janssens and I. Prigogine, Physica \textbf{XVI}, 895 (1950); J.S. Rowlinson
and C.F. Curtis, J. Chem. Phys. \textbf{19}, 1519 (1951); J.M.H. Leelt and
R.P. Hurst, J. Chem. Phys. \textbf{32}, 96 (1960).

\bibitem {NoteEntropy}It should be emphasized that the way it has been
introduced, $W(E,V)dE/\epsilon_{0}$ does not either have to be an integer or
unique because of the introduction of $v_{0}$ and $\epsilon_{0}$. This is a
common problem in classical statistical mechanics and one must go a quantum
description to make the number of microstates to be an integer and unique.
Despite this, the communal entropy in (\ref{Communal_S}) is expected to be
non-negative even in classical thermodynamics if it is associated with the
number of basins or with deconfinement; see Sect. \ref{Lattice_Model}.

\bibitem {note0}For a macroscopic system, we will always neglect terms that do
not grow exponentially fast, as they are not relevant in the thermodynamic
limit $N\rightarrow\infty$.

\bibitem {GujFedor}P.D.\ Gujrati and F. Semerianov, cond-mat/0412759.\ 

\bibitem {Hill}T.L. Hill, Statistical Mechanics, Dover (1956), p. 355.

\bibitem {NernstPostulate}We will assume that the entropy $S(T)$ of stationary
states (EQS's or SMSs) satisfy $TS(T)\rightarrow0$ as the temperature
$T\rightarrow0$. This is a much weaker condition than the conventional
Nernst-Planck postulate $S(T)\rightarrow0$ for EQS's, but is consistent with
all the consequences of the latter. Our version is also applicable to SMSs,
for which the entropy need not vanish at absolute zero.\cite[Sect. 64]{Landau}

\bibitem {Huang}K. Huang, \textit{Statistical Mechanics}, (second edition),
John Wiley, New York (1963).

\bibitem {Ruelle}D. Ruelle, Physica (Utrecht) \textbf{113A}, 619 (1982).

\bibitem {CorsiThesis}A. Corsi, Ph.D. Dissertation, University of\ Akron (2004).

\bibitem {concave}A concave function $f(x)$ is a function that always lies
above the line connecting $f\left(  x_{1}\right)  $ and $f\left(
x_{2}\right)  $ over any of its interval $\left[  x_{1},x_{2}\right]  $.

\bibitem {Gujrati-Glass-Defective-Crystal}P.D. Gujrati, arXiv:0708.2075.

\bibitem {Note1}Of course, we are considering a nono-local defect whose size,
determined by the number of sites contained in it, is strictly less than $N$. 

\bibitem {Lindemann}F.A. Lindemann, Phys. Z., \textbf{11}, 609 (1910).

\bibitem {Jin}Z. H. Jin, P. Gumbsch, K. Lu, and E. Ma, Phys. Rev. Lett., 87,
055703, (2001).

\bibitem {Zallen}R. Zallen, \textit{The Physics of Amorphous Solids}, John
Wiley and Sons, New York (1983).

\bibitem {Flory}P.J. Flory, \textit{The Principles of Polymer Chemistry},
Cornell University Press, Ithaca (1953).\ 

\bibitem {deGennes}P.G. deGennes, \textit{Scaling Concepts in Polymer Physics}
Cornell University Press, Ithaca (1979).\ 

\bibitem {Novikov}V.N. Novikov, E. R\"{o}ssler, V.K. Malinovsky and N.V.
Surovtsev, Europhys. Lett. \textbf{35}, 289 (1996).

\bibitem {deGennesGlass}P.G. deGennes, C.R. Physique, \textbf{3}, 1263 (2002).

\bibitem {GujResidualEntropy}P.D. Gujrati, arXiv:0908.1075.

\bibitem {Kubo}R. Kubo, \textit{Statistical Mechanics}, North-Holland, 1981.

\bibitem {Berker}R. Netz and N. Berker, Phys. Rev. Lett. \textbf{66}, 377 (1991).

\bibitem {GujPRL}P. D. Gujrati, Phys. Rev. Lett. \textbf{74}, 809 (1995).

\bibitem {Cerena}Cerena Uttal and P.D. Gujrati, unpublished.

\bibitem {Keith}Keith P. Pelletier and P.D. Gujrati, unpublished.

\bibitem {Domb}C. Domb in \textit{Phase Transitions and Critical Phenomena},
Vol. 3, edited by C. Domb and M.S. Green, Academic Press (1974).
\end{thebibliography}
\end{document}